\documentclass[twocolumn]{aastex62}

\newcommand{\chn}{{\it Chandra}}
\usepackage{blindtext}
\usepackage[colorinlistoftodos]{todonotes}
\usepackage{lineno}
\usepackage{amsmath}
\usepackage{MnSymbol}
\usepackage{graphicx}
\usepackage{appendix}
\usepackage{multirow}
\usepackage{url}
\usepackage{wasysym}
\begin{document}
%-----------------------------------------------------------------------------------------------

\title{Powerful yet lonely: Is 3C\,297 a high-redshift fossil group$?$}

\author[0000-0001-8382-3229]{Valentina Missaglia} 
\affiliation{Dipartimento di Fisica, Universit\`a degli Studi di Torino, via Pietro Giuria 1, I-10125 Torino, Italy}
\affiliation{INFN-Istituto Nazionale di Fisica Nucleare, Sezione di Torino, I-10125 Torino, Italy.}
\affiliation{INAF-Osservatorio Astrofisico di Torino, via Osservatorio 20, I-10025 Pino Torinese, Italy}

\author{Juan P.\ Madrid}
\affiliation{Departamento de F\'isica y Astronom\'ia, La Universidad de Tejas en el Valle del R\'io Grande, Brownsville, TX 78520, USA}

\author[0000-0003-2568-9994]{Mischa Schirmer}
\affiliation{Max-Planck-Institut f\"ur Astronomie, K\"onigstuhl 17, D-69117 Heidelberg, Germany}

\author[0000-0002-1704-9850]{Francesco Massaro} 
\affiliation{Dipartimento di Fisica, Universit\`a degli Studi di Torino, via Pietro Giuria 1, I-10125 Torino, Italy}
\affiliation{INFN-Istituto Nazionale di Fisica Nucleare, Sezione di Torino, I-10125 Torino, Italy.}
\affiliation{INAF-Osservatorio Astrofisico di Torino, via Osservatorio 20, I-10025 Pino Torinese, Italy}

\author{Alberto Rodr\'iguez-Ardila}
\affiliation{LNA/MCTIC, Rua dos Estados Unidos, 154. Bairro das Na\c{c}\~oes, Itajub\'a, MG B-37501-591, Brazil}
\affiliation{Divis\~ao de Astrof\'isica, INPE, Avenida dos Astronautas 1758, S\~ao Jos\'e dos Campos, B-12227-010 SP, Brazil}

\author{Carlos J. Donzelli}
\affiliation{Instituto de Astronom\'ia Te\'orica y Experimental 
IATE, CONICET -- Observatorio Astron\'omico, Universidad Nacional de C\'ordoba, Laprida 854,
X5000BGR, C\'ordoba, Argentina}

\author{Martell Valencia}
\affiliation{Departamento de F\'isica y Astronom\'ia, La Universidad de Tejas en el Valle del R\'io Grande, Brownsville, TX 78520, USA}

\author[0000-0002-5646-2410]{Alessandro Paggi}
\affiliation{Dipartimento di Fisica, Universit\`a degli Studi di Torino, via Pietro Giuria 1, I-10125 Torino, Italy}
\affiliation{INFN-Istituto Nazionale di Fisica Nucleare, Sezione di Torino, I-10125 Torino, Italy.}
\affiliation{INAF-Osservatorio Astrofisico di Torino, via Osservatorio 20, I-10025 Pino Torinese, Italy}

\author[0000-0002-0765-0511]{Ralph P.\ Kraft}
\affiliation{Center for Astrophysics,  Harvard \& Smithsonian, 60 Garden Street, Cambridge, MA 02138, USA}

\author[0000-0003-1619-3479]{Chiara Stuardi} 
\affiliation{Dipartimento di Fisica e Astronomia, Universit\'a di Bologna, via Gobetti 93/2, I-40129 Bologna, Italy}
\affiliation{INAF - Istituto di Radioastronomia di Bologna, Via Gobetti 101, I-40129 Bologna, Italy}

\author[0000-0003-1809-2364]{Belinda J. Wilkes}
\affiliation{Center for Astrophysics,  Harvard \& Smithsonian, 60 Garden Street, Cambridge, MA 02138, USA}
\affiliation{School of Physics, University of Bristol, Bristol, UK}

\correspondingauthor{Juan P.\ Madrid} 
\email{juan.madrid@utrgv.edu}

%-----------------------------------------------------------------------------------------------
                         
\begin{abstract}

The environment of the high-redshift ($z=1.408$), powerful radio-loud galaxy 3C\,297 has several distinctive features of a galaxy cluster. Among them, a characteristic halo of hot gas revealed by Chandra X-ray observations. In addition, a radio map
obtained with the Very Large Array (VLA) shows a bright hotspot in the northwestern direction, created by the interaction of the AGN jet arising from 3C\,297 with its environment. In the X-ray images, emission cospatial with the northwestern radio lobe is detected, and peaks at the position of the radio hotspot. The extended, complex X-ray emission observed with our new Chandra data is largely unrelated to its radio structure. Despite having attributes of a galaxy cluster, no companion galaxies have been identified from 39 new spectra of neighboring targets of 3C\,297 obtained with the Gemini Multi-Object Spectrograph. None of the 19 galaxies for which a redshift was determined lies at the same distance as 3C\,297. The optical spectral analysis of the new Gemini spectrum of 3C\,297 reveals an isolated Type-II radio-loud AGN. We also detected line broadening in [O\,{\sc ii}]~$\lambda$3728 with a FWHM about 1700 km s$^{-1}$ and possible line shifts of up to 500-600 km s$^{-1}$. We postulate that the host galaxy of 3C\,297 is a fossil group, in which most of the stellar mass has merged into a single object, leaving behind an X-ray halo.

\end{abstract}


%-----------------------------------------------------------------------------------------------

%-----------------------------------------------------------------------------------------------
\section{Introduction}
%-----------------------------------------------------------------------------------------------

The sample of powerful radio sources in the Third Cambridge Revised Catalog \citep[3CR;][]{bennett1962a,bennett1962b} has served for decades as the foundation to investigate the nature and evolution of radio-loud active galactic nuclei \citep[AGN; see][for a review]{merloni2013}. There is consensus that strong extragalactic radio sources are at the core of massive elliptical galaxies, powered by supermassive black holes (SMBH) \citep[][]{rees1984,begelman1984}. Radio-loud AGN launch bright, powerful plasma jets \citep[e.g.][]{salpeter1964, lynden1969,padovani2017,blandford2019} that typically extend over scales much larger than the host galaxy itself, interacting with -- and sometimes shaping -- the surrounding intracluster medium (ICM). This phenomenon, known as feedback \citep[see e.g.,][]{fabian2012}, can also ignite or quench star formation. An iconic example is M87 \citep[3C\,274;][]{owen2000, churazov2001,forman2007,dega2012,perley2017}, whose SMBH has recently been directly imaged \citep{eventhorizon2019}.

Some of these powerful radio sources are typically the most luminous members of the galaxy cluster in which they reside, being therefore called Brightest Cluster Galaxies \citep[BCGs;][]{best2007}, usually fed by ``cooling flows" \citep[e.g.][]{fabian2012,bykov2015}. 

At high redshift ($z>1$), some of these radio sources act as beacons indicating the potential presence of galaxy clusters whose members have not been yet identified \citep[e.g.][]{wing2011,paterno2017}. In particular, at redshift $\sim$ 2, it is possible to study the epoch of galaxy cluster formation, to understand how the ICM properties evolve and how this can affect cluster formation and evolution \citep[see e.g.][and references therein]{rosati2002}. 

The work presented here focuses on 3C\,297, a radio-loud quasar at $z=1.406$ \citep{spinrad1985}, and suggests the presence of a massive cluster in formation. \citet{spinrad1985} emphasized the high redshift of 3C\,297,
and reported the detection of [Mg\,{\sc ii}], [Ne\,{\sc iv}], and [C\,{\sc iii}] lines
in its spectrum. \citet{jackson1997} took a near-infrared spectrum of 3C\,297 between
1.05 and 1.25 $\mu$m, detecting a prominent [O\,{\sc iii}]~$\lambda$5007
line redshifted to 1.2053~$\mu$m corresponding to a redshift measurement $z=1.407$.

This source has a large suite of recent multiwavelength observations. 
\citet{chiaberge2015} classify 3C\,297 as a merger, using Hubble Space Telescope (\textit{HST}) observations, because of the presence of a double core.
\citet{hilbert2016} reported earlier radio, optical, and infrared observations of 3C\,297. A Very Large Array (VLA) map shows a strong, straight lobe of radio emission stretching
from the host galaxy to more than 5\arcsec\ ($\sim$~43 kpc) to the northwest. In addition, 3C\,297 has diffuse radio emission to the south of the host galaxy. The southern radio emission includes two knots of optical emission, whose relative position is uncertain given the absence of a radio core to align the radio and the optical/IR. From observations obtained with the UVIS channel of \textit{HST}'s Wide Field Camera 3 through the F606W filter and with the IR channel through the F140W filter, 3C\,297 appears as a compact but resolved core with diffuse emission, with an arc 2\arcsec\ north of the core and a highly elongated source 1\arcsec\ to the south. This highly elongated source is suggestive of a recent/ongoing merger as already claimed by \citet{chiaberge2015}. 

\cite{kotyla2016} analyzed \textit{HST} data of 21 high-$z$ sources of the 3C catalog and  noticed that 3C\,297 lacks an associated galaxy cluster or group. 

In the Chandra Snapshot Survey of 3C sources \citep{massaro2015,massaro2018,stuardi2018}, 3C\,297 shows a complex morphology of its X-ray halo as the likely result of an interaction with the radio jet \citep[``feedback", see e.g.][]{fabian2012,kraft2012}.

The combination of multiobject spectroscopy and Chandra data proved to be an efficient method to find that another 3C source belongs to a galaxy cluster. In \cite{madrid2018} we used Gemini spectra and Chandra images to identify a previously unknown cluster of galaxies around 3C\,17, at $z=0.22$. 3C\,17 has been of particular interest because its radio jet is dramatically bent \citep{morganti1999,massaro2009,massaro2010}, a sign of strong interaction with the ICM. 

While the presence of hot gas and the bent radio jet of 3C\,297 are typical for galaxy clusters, no galaxies at the same redshift are known in the literature around 3C\,297 within $\sim$2 Mpc. These findings prompted us to do follow-up observations of 3C\,297
in an effort to find potential companions. 

In this paper, we present new Gemini multiobject spectroscopy of field galaxies in the vicinity of 3C\,297, to search for a potential galaxy group or cluster members, and new Chandra observations performed to distinguish between the X-ray emission arising from the ICM and that related to the large-scale radio structure.

In Section 2, we describe the Chandra data reduction and analysis. In Section 3, we describe our new  Gemini observations. Section 4 is devoted to the analysis of Gemini optical spectra. In Section 5, we discuss the possibility that 3C 297 may be the central galaxy of a fossil group. A brief summary and our conclusions are given in Section 6. Throughout the paper we assume a $\Lambda$CDM cosmology with $H_0$=69.6\,km\,s$^{-1}$\,Mpc$^{-1}$, $\Omega_{\text{M}}$=0.286, and $\Omega_{\Lambda}$=0.714 (\citealt{bennett2014}).

\begin{figure}
  \epsscale{1.2}
  \plotone{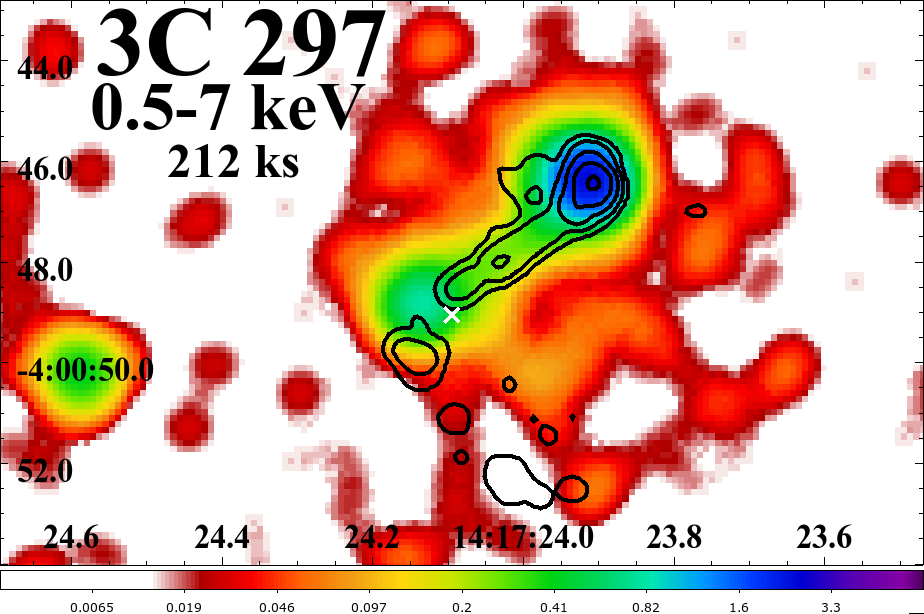}
  \caption{\chn\ image of 3C\,297, obtained after merging the available observations, for a total of 212 ks. The image has been filtered in the energy range 0.5-7 keV, binned with a pixel size of 0.123\arcsec/pixel and smoothed with a 3.44\arcsec\ Gaussian kernel. VLA 8.4 GHz (restoring beam 0.357\arcsec~$\times$~0.233\arcsec) black radio contours are drawn starting at 0.08~$\mu$Jy beam$^{-1}$, increasing by factors of six. The position of the radio core, derived from the spectral index map, is marked with a white cross.\\
  \label{smooth}}
\end{figure}

%-------------------------------------------------------------------------------------------
\section{Chandra observations, data reduction and analysis}
\label{xray}
%-----------------------------------------------------------------------------------------------
3C\,297 was observed by \chn\ in 2016 as part of the 3C Snapshot Survey \citep[see][for a recent review]{stuardi2018} for a total exposure of $\sim$ 12 ks (OBSID: 18103).

New Chandra ACIS-S data in VERY FAINT mode (P.I. Missaglia) were obtained between 2021 April and 2022 April (see Figure~\ref{smooth}). The awarded observing time has been divided into 10 observations, for a total nominal exposure time of $\sim$ 200 ks (OBSIDs: 23829; 24352, 24353, 24355, 24356, 24357, 25000, 25001, 25004, 25023).
\chn\ data reduction was carried out using the Chandra Interactive Analysis of Observations \citep[CIAO v4.13][]{ciao} following standard procedures and threads\footnote{http://cxc.harvard.edu/ciao/threads/}, and the Chandra Calibration Database (CALDB) version 4.9.6. 
Level 2 event files were created using the CIAO task \texttt{chandra\_repro}. We removed flares from lightcurves\footnote{https://cxc.harvard.edu/ciao/ahelp/deflare.html}, and astrometrically registered all the images using the VLA 8.4 GHz radio observation, smoothing the X-ray observations to find the brightest pixel corresponding to the northern hotspot as in the VLA image. This association is permissible within the astrometric uncertainties of both data sets.
The VLA data we use was obtained in the A configuration (maximum baseline 36 km) on 1990 May 10 at 8.44\,GHz.

\citet{stuardi2018} suggested a tentative position of the radio core, based on the results of the spectral index map. However, to register the X-ray images, we used the northern hotspot because it is a bright feature, easy to recognize.
As a last step, we merged all the observations\footnote{\url{https://cxc.harvard.edu/ciao/ahelp/reproject\_obs.html}} for the spectral analysis of the most interesting features.

\begin{figure}
  \epsscale{1.2}
  \plotone{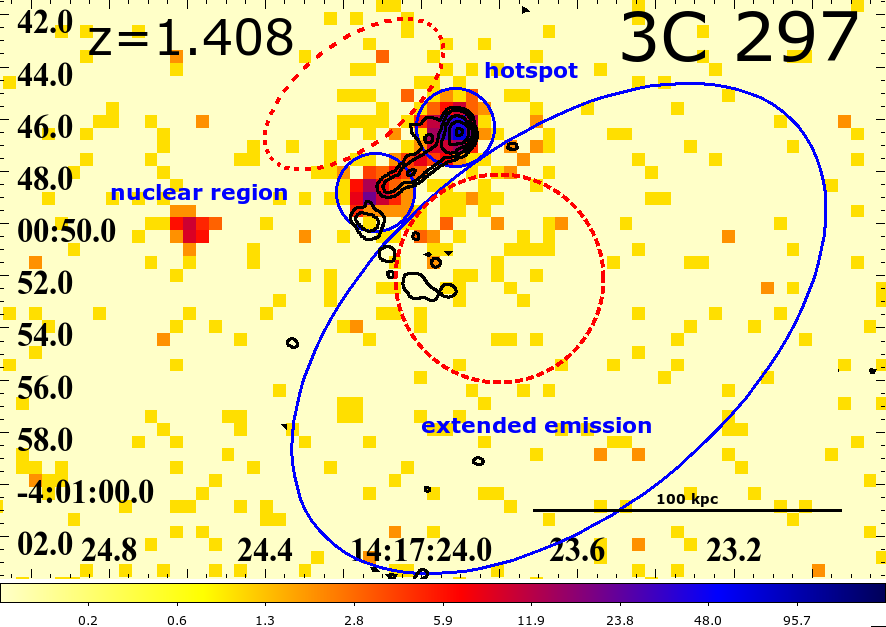}
  \caption{Unbinned, unsmoothed \chn\ merged image of 3C\,297 in the energy band 0.3-7 keV. The three regions in which spectral analyses have been performed are highlighted in the blue regions (two circles and one ellipse). We detected 159 net counts in the extended emission region, 115 net counts in the nuclear region and 315 net counts in the hotspot region. The red dashed circle and ellipse mark the regions in which we observe an X-ray counts excess, due to the outflow (see Section~\ref{outflow} for more details). \\
  \label{regions}}
\end{figure}

\begin{figure}
  \epsscale{1.2}
  \plotone{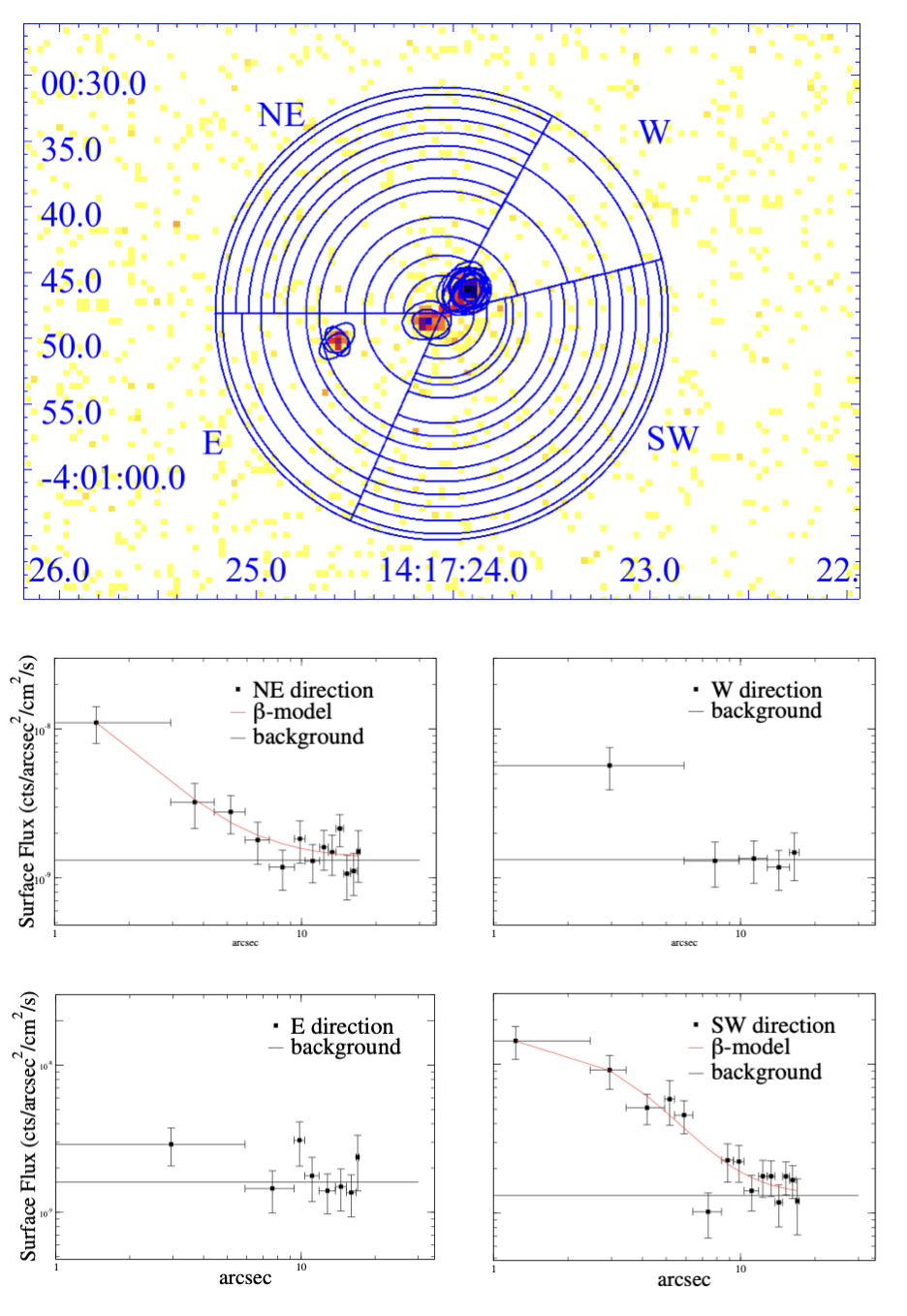}
  \caption{\textit{Top Panel}: Merged \chn\ X-ray image (0.5-7 keV) showing the annuli used to extract surface brightness profiles in four sectors. Blue ellipses represent the point sources excluded from each observation before merging the separate pointings. The source in the eastern region is unrelated. \textit{Bottom Panel}: Surface brightness profiles for the regions highlighted in the top panel. In the western and eastern directions the emission cannot be fitted with a $\beta$-model.
  \label{expcorr}}
\end{figure}

The spectral analysis focused on three distinct areas with sufficient counts: the northern hotspot, the extended region, and the nucleus (see Figure~\ref{regions}). Before this, we extracted the surface brightness profiles in four directions (see Figure~\ref{expcorr}) choosing sections with an outer radius of 30\arcsec\ and a signal-to-noise ratio of 3 in each radial bin. We then extracted the spectra with the \texttt{specextract}\footnote{https://cxc.cfa.harvard.edu/ciao/ahelp/specextract.html} script and performed the fitting with Sherpa (\citealt{freeman2001}).

We extracted the spectrum of the hotspot within a circular aperture of $\sim$ 1.5\arcsec\ diameter in the reprojected observation, and we extracted the spectrum merging the 11 available observations (snapshot plus new observations). Adopting a model with Galactic absorption and a power law (xstbabs.absgal*xspowerlaw.pl) setting the Galactic absorption to the value of $3.49\times10^{20}$ cm$^{-2}$, as reported in \citet{hi4pi2016}, we obtained a photon index $\Gamma=1.8\pm0.16$ and an X-ray luminosity of the hotspot L$_X=2.82^{+0.57}_{-0.46}\times10^{44}$ erg s$^{-1}$ between 0.5 and 7 keV.

We performed the spectral analysis of the extended emission in the southwestern direction.
We extracted spectra from an elliptical region (see Fig.~\ref{regions}) with semi-major and minor axes of 12\arcsec\ and 7\arcsec, respectively. The axes were selected based on the surface brightness profile in the southwestern direction.

To investigate if this emission is non thermal, due to inverse Compton scattering (IC) of non-thermal radio-emitting electrons on cosmic microwave background (CMB) photons \citep[IC/CMB;][]{felten1969,cooke1978,harris1979}, or thermal, and therefore related to the ICM, we adopted two different models for the extended emission: (1) Galactic absorption plus a power law (xswabs.absgal*xspowerlaw.pl), and (2) Galactic absorption plus a thermal model (xswabs.absgal*xsapec.therm). 

As reported in the literature \citep[see e.g.][and references therein]{massaro2011,ghisellini2015}, the magnetic fields in the lobes of radio galaxies are relatively low, of the order of tens of microgauss, therefore radiating synchrotron emission inefficiently.

At high $z$ the CMB energy density dominates the radio-lobe magnetic field energy density. Therefore, as described by \citet[][and references therein; see also \citealt{volonteri2011}]{ghisellini2014a, ghisellini2014b}, at high $z$ the high-energy electron population is depleted due to scattering on the CMB photons, which serves to reduce the radio brightness of the lobes (CMB quenching). At the same time, the increased CMB energy density, enhances the IC/CMB X-ray emission from the lobes (\citealt{rees1968}). 

\citet{yuan2003} presented a \chn\ analysis of the extended X-ray emission around the high-redshift ($z$=4.301) quasar GB1508+5714. This emission is well described as IC/CMB, and the lack of an obvious detection of radio emission from the extended component could be a consequence of Compton losses on the electron population, or of a low magnetic field. It is also argued that extended X-ray emission produced by IC scattering may be common around high-redshift radio galaxies and quasars, demonstrating that significant power is injected into their surrounding by powerful jets.

In the nonthermal scenario (X-ray diffuse emission due to IC/CMB in the lobes), we obtained a photon index equal to $\Gamma=2.72^{+0.94}_{-0.77}$ and a luminosity of L$_X=6.6^{+3.3}_{-2.5}\times10^{43}$ erg s$^{-1}$ (see details in Table~\ref{tab:spec}). 

Given that the diffuse X-rays are not associated with the radio structure, we also had to consider the thermal scenario (X-ray emission from the ICM), even if in the case of cluster emission we should expect a spherically symmetric emission. In this case, adopting a (xswabs.absgal*xsapec.therm) model, we could only put an upper bound on the temperature of 6 keV. This translates to an X-ray luminosity between 0.5 and 7 keV of L$_X=4.5^{+2.0}_{-1.7}\times10^{43}$ erg s$^{-1}$ ($\chi^{2}$(dof)=0.72(6)).
To disentangle the two scenarios, we would need low-frequency radio data that could trace the extension of the southern lobe and allow us to estimate the IC/CMB contribution to the X-ray emission from the lobe region.

Finally, we extracted the spectrum of the nuclear region, i.e.\ the region that in the optical corresponds to the host galaxy of 3C\,297 (the position of the central AGN identified with the radio spectral index map as reported in \citet{stuardi2018} that is marked with a white cross in Fig.~\ref{smooth}).
We adopted a model with Galactic and intrinsic absorption, plus a power law (xswabs.absgal*xszwabs.absint*xspowerlaw.pl). We fitted the spectrum in three different configurations: (1) with photon index, normalization, and intrinsic absorption as free parameters; (2) with photon index equals 1.8 and free intrinsic absorption and normalization, and (3) with no intrinsic absorption. Results of the spectral fitting for the three components are summarized in Table~\ref{tab:spec}.
The nucleus has an X-ray luminosity of L$_X=1.15^{+0.53}_{-0.40}\times10^{44}$ erg s$^{-1}$ between 0.5-7 keV.

\begin{table*}
\centering
\begin{tabular}{ |c c c c c c| }
 \hline
 \multicolumn{6}{|c|}{Results of the X-Ray spectral fitting} \\
 \hline
\hline
  & \multicolumn{5}{c|}{Power-Law} \\
 \cline{2-6}
  Region & N$_{H,int}$ & $\Gamma$ & norm & $\chi^{2}_{\nu}$(dof) & L$_{X(0.5-7\text{keV})}$ \\
  %\cline{2-6}
  & (10$^{22}$ cm$^{-2}$) & & (10$^{-6}$ cm$^{-5}$) & & (10$^{44}$ erg s$^{-1}$) \\
  \hline
 Northern Hotspot  & -  & $1.87\pm0.15$ & 4.87$^{+0.65}_{-0.22}$ & 0.68(13) & 2.82$^{+0.57}_{-0.46}$ \\
 \hline
 X-ray Halo & - & 2.72$^{+0.94}_{-0.77}$ & 1.50$^{+0.58}_{-0.54}$ & 0.63(6) & 0.66$^{+0.33}_{-0.25}$\\
 \hline
 \multirow{3}{*}{Nucleus} & 1.95$\times10^{-4}$ & 1.44$^{+0.51}_{-0.31}$ & 1.42$^{+1.27}_{-0.37}$ & 0.42(2) & \\
 & $>$4.7 & 1.8* & 2.16$^{+0.51}_{-0.45}$ & 0.48(3) &1.15$^{+0.53}_{-0.40}$\\
 & - & 1.44$^{+0.33}_{-0.31}$ & 1.42$^{+0.46}_{-0.39}$ & 0.28(5) &\\
 \hline \hline
 & \multicolumn{5}{c|}{Thermal}\\
 \cline{2-6}
 Region & N$_{H,int}$ & kT & norm & $\chi^{2}_{\nu}$(dof) & L$_{X}(0.5-7 keV)$ \\
  %\cline{2-5}
 & (10$^{22}$ cm$^{-2}$) & (keV) & (10$^{-6}$ cm$^{-5}$) & & (10$^{43}$ erg s$^{-1}$)\\
  \hline
 X-ray Halo & - & $<$6 & 25$^{+34}_{-13}$ & 0.72(6) & 4.5$^{+2.0}_{-1.7}$\\
 \hline
 
\end{tabular}
\caption{Results of the spectral fitting of the X-Ray components: northern hotspot, extended emission and nucleus (all highlighted in Fig.~\ref{regions}). * marks fixed values in the fit.}
\label{tab:spec}
%\end{centering}
\end{table*}

%-----------------------------------------------------------------------------------------------
\section{Gemini observations}
%-----------------------------------------------------------------------------------------------

%--------------------------------------------------------------------------------------------
\begin{figure*}
\epsscale{0.8}
\plotone{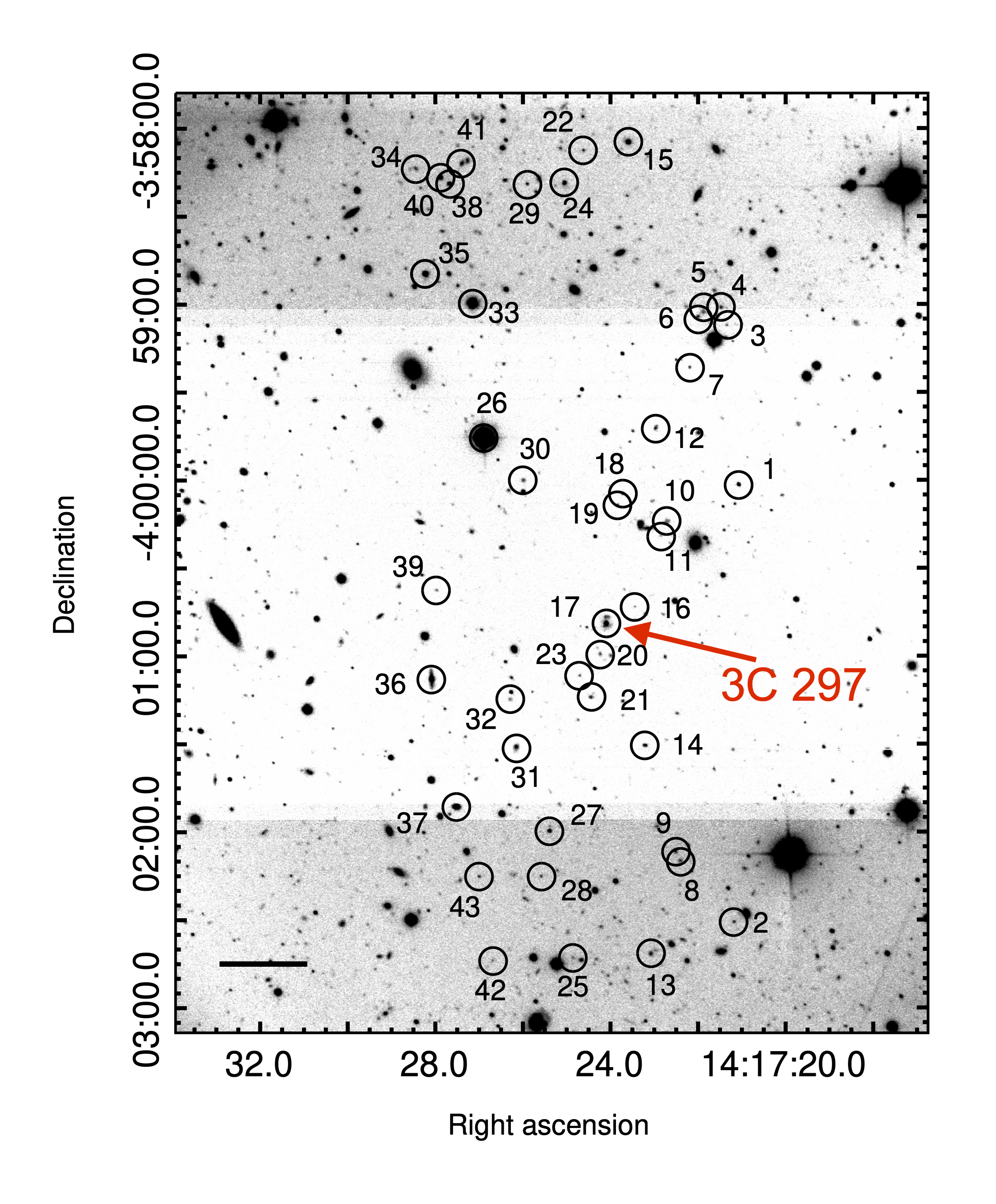}
\caption{Section of the GMOS preimage of the 3C\,297 field. This $i$-band image has a total exposure time of 1080 s. Numbers correspond to the slit ID. The radio galaxy 3C\,297 is  close to the center of the image, identified with slit number 17. The bar on the lower left represents 30$\arcsec$. The different background levels are a due to limited processing functionality by the Gemini pipeline that automatically produces the preimage. North is up and east is left.\\
\label{preimage}}
\end{figure*}
%-------------------------------------------------------------------------------------------

\subsection{Preimage and Target selection}

For  field spectroscopy, we used the Gemini Multi-Object Spectrograph
\citep[GMOS;][]{hook2004}. The $i$-band preimage for the mask design was
obtained on 2020 February 18 through program GS-2019B-FT-211 (PI: Madrid)
with GMOS at Gemini South. The integration time was 1080\,s, taken in
clear and 0.75\arcsec\ seeing conditions. 

In our GMOS $i$-band preimage (see Figure~\ref{preimage}), 3C\,297 has an $i$-band magnitude of 20.7\,AB mag. Lacking photometric redshifts in the field, our initial target selection was based on the $i$-band photometry of galaxies in this image.
If there was a galaxy cluster around 3C\,297, then most members would be considerably fainter than the BCG -- in the case of a fossil group, by at least 2.5\,mag within the virial radius \citep{dariush2007}. We therefore selected our high-priority target sample from the $21.8<i<23.0$\,AB mag interval, corresponding to $0.2$--$3.0\times10^{10}$\,$L_{\odot}$ if at $z\sim1.4$. The bright cutoff was chosen to reduce the contamination by foreground sources. For targets below the faint cutoff it would be increasingly difficult to determine a spectroscopic redshift, in particular if they reside in the redshift desert between $z\sim1.8$--$2.5$, where redshifted emission lines are absent in the optical wavelength range. Galaxies with $i=23.0$--$24.0$\,mag comprised a low-priority  sample that the mask-making algorithm could use to fill any remaining spaces in the mask design.

After the Gemini mask-making software automatically resolved slit-placement conflicts (spatial or spectral overlaps), 40 galaxies -- or about 50\% of the targets -- remained in the mask design. Science targets had slitlets of 1\arcsec. While this sample of 40 targets is necessarily incomplete, it is still representative, enabling us to detect a potential cluster population in the spatial vicinity of 3C\,297. In addition to these galaxies, we had two alignment stars in wider slits, and one reference star and three galaxies of similar brightness as 3C\,297 in narrow science slits, to guide the data processing and judge data quality. 3C\,297 was also included, near the center of the mask (see Table~\ref{tbl-1} and Figure~\ref{preimage}).

\subsection{Band-shuffling mask design}

We used GMOS in its band-shuffling mode, which allows for very high slit
density in a fairly compact area. Specifically, one-third of the detector area or
$5.5^\prime\times1.8^\prime$ are available for the slit placement,
corresponding to a projected physical area of $2.8\times0.9$\,Mpc at
$z=1.408$.

During the observations, science targets are observed  for 60\,s, after which the collected charges are moved (``shuffled'') to an unilluminated detector area, and the telescope makes a small offset. During the next 60\,s, at the offset position, the slits collect sky signal only. Then the \textit{target} charges are shuffled back under the slit, and the sky charges are moved onto a different storage area. This is repeated for a total of 600\,s in a single exposure, for an effective on-source integration time of 300\,s. Very accurate sky subtraction and fringe correction becomes possible at longer wavelengths dominated by airglow, with just a single readout, and redshifts can be secured for fainter galaxies. For details about this mode see \citet{glaze2001}.

%-----------------------------------------------------------------------

\subsection{Instrumental setup}

Our instrumental setup is optimized to identify galaxies at a redshift of $z\sim 1.4$. Many characteristic emission lines of active and star-forming galaxies, such as [MgII]$\lambda$2799, [OII]$\lambda$3727, [NeIII]$\lambda$3869, and H$\delta$, would be redshifted to the 670--990\,nm range. This interval then also includes the characteristic Balmer break and the Ca H+K absorption lines of older stellar populations in elliptical galaxies. Our instrumental setup for GMOS, therefore, used the R400 grating with the OG515 order-sorting filter and a central wavelength of 790\,nm. The associated spectral resolution for a slit width of 1\arcsec\ is $R=960$. The main observable wavelength interval with this setup was $510$--$1050$\,nm, with negligible second-order overlap above  1020\,nm. Depending on how close a source was located to the edge of the masking area, the spectral range was truncated by the detector edges either at the blue or the red end. Typically, we covered a 450\,nm wide interval within $510$--$1050$\,nm for each source.

\subsection{Observations}

One month after our original mask was cut and tested, Gemini suspended its science operations due to the pandemic, preventing our program to proceed with the spectral observations. By the time Gemini South reopened, the GMOS-S detector was experiencing considerable charge-transfer efficiency issues that were prohibitive for our band-shuffling observations.

The mask was thus redesigned for identical band-shuffling observations with GMOS-N at Gemini North, using the GMOS-S preimage. Spectra were obtained on 2021 April 13 and 15 under program GN-2021A-FT-203 (PI: Madrid), with an effective on-source integration time of $11\times300 = 3300$\,s, and an image seeing of $0.5\arcsec$--$0.8$\arcsec. We binned the detector 2$\times$2 (in the spatial and spectral dimensions), and observed two central wavelength settings of 790\,nm and 800\,nm to cover the detector gaps. Calibration data consisted of biases, lamp and twilight flats, and CuAr arc spectra for wavelength calibration.

\subsection{Data reduction}

The data reduction was carried out following a sequence of {\sc Pyraf} Gemini/GMOS routines. Biases and flat-fields were reduced using the tasks {\sc gbias} and {\sc gsflat}. Spectra for science targets and the CuAr lamps were processed with the {\sc gsreduce} task. The {\sc gswavelength} routine 
was used to derive the wavelength calibration from the CuAr frames that were applied  to
the science spectra with {\sc gstransform}. Sky subtraction was performed with {\sc gnsskysubtract} and final 1-D spectra extraction was carried out with the routine {\sc gsextract}.

Flux calibration was achieved using GMOS longslit observations of the spectroscopic standard
star \textit{Feige66} made during the observing run. All spectra were flux calibrated and
extinction corrected with the {\sc gscalibrate} task. Lastly, we combine all 1D spectra 
into a single spectrum for each target using {\sc gemcombine}.

%-----------------------------------------------------------------------------------------------
%---------------------------------------------------------------------
%--Tabla 1------------------------------------------------------------
\begin{center}
\begin{deluxetable}{lcccccl} 
\tabletypesize{\scriptsize} 
\tablecaption{Targets of the GMOS multi-slit mask \label{tbl-1}} 
\tablewidth{0pt}
\tablehead{
\colhead{Slit} & \colhead{R.A.} & \colhead{Dec.} & \colhead{m$_{i}$} & \colhead{z}}
\startdata
1  &  14:17:21.076 &  -4:00:01.45 &     21.45 & ---    \\
2  &  14:17:21.177 &  -4:02:30.65 &     23.37 & ---    \\
3  &  14:17:21.344 &  -3:59:07.30 &     23.91 & $\medstar$  \\
4  &  14:17:21.475 &  -3:59:01.19 &     22.92 & ---    \\
5  &  14:17:21.871 &  -3:59:02.67 &     22.96 & ---    \\
6  &  14:17:22.008 &  -3:59:04.24 &     23.01 & 1.156  \\
7  &  14:17:22.193 &  -3:59:21.65 &     23.26 & 0.633  \\
8  &  14:17:22.400 &  -4:02:09.94 &     23.09 & 1.263  \\
9  &  14:17:22.525 &  -4:02:06.74 &     22.87 & 1.163  \\
10 &  14:17:22.737 &  -4:00:14.41 &     22.68 & 1.029  \\ 
11 &  14:17:22.855 &  -4:00:19.08 &     23.81 & ---    \\
12 &  14:17:22.977 &  -3:59:42.29 &     22.51 & 1.523  \\
13 &  14:17:23.077 &  -4:02:41.65 &     22.63 & ---    \\
14 &  14:17:23.204 &  -4:01:30.48 &     22.28 & ---    \\       
15 &  14:17:23.588 &  -3:58:04.49 &     21.23 & 0.215  \\
16 &  14:17:23.455 &  -4:00:43.17 &     24.25 & ---    \\       
\textbf{17} &  \textbf{14:17:24.102} &  \textbf{-4:00:48.93} &     \textbf{20.95} & \textbf{1.408}  \\
18 &  14:17:23.719 &  -4:00:04.45 &     22.74 & ---    \\
19 &  14:17:23.839 &  -4:00:08.48 &     23.22 & 1.137  \\
20 &  14:17:24.232 &  -4:00:59.38 &     23.21 & ---    \\
21 &  14:17:24.447 &  -4:01:14.20 &     22.89 & ---    \\
22 &  14:17:24.621 &  -3:58:07.42 &     23.44 & 1.343  \\
23 &  14:17:24.702 &  -4:01:06.75 &     23.57 & 1.577  \\
24 &  14:17:25.052 &  -3:58:18.53 &     21.91 & ---    \\
25 &  14:17:24.861 &  -4:02:43.03 &     23.80 & ---    \\
26 &  14:17:26.898 &  -3:59:45.86 &     15.12 & $\medstar$ \\
27 &  14:17:25.387 &  -4:01:59.64 &     21.88 & ---    \\
28 &  14:17:25.583 &  -4:02:15.24 &     23.56 & 0.732  \\
29 &  14:17:25.901 &  -3:58:18.92 &     23.10 & 2.577  \\
30 &  14:17:25.989 &  -4:00:00.01 &     22.87 & ---    \\
31 &  14:17:26.160 &  -4:01:30.80 &     21.69 & 0.211  \\
32 &  14:17:26.292 &  -4:01:14.94 &     23.35 & 1.181  \\
33 &  14:17:27.150 &  -3:58:59.79 &     17.44 & $\medstar$  \\
34 &  14:17:28.444 &  -3:58:13.55 &     23.12 & ---    \\
35 &  14:17:28.238 &  -3:58:49.56 &     20.34 & ---    \\
36 &  14:17:28.087 &  -4:01:07.61 &     20.20 & ---    \\
37 &  14:17:27.513 &  -4:01:51.40 &     20.38 & 0.244  \\
38 &  14:17:27.675 &  -3:58:18.69 &     21.72 & 0.295  \\
39 &  14:17:27.965 &  -4:00:37.53 &     23.32 & ---    \\
40 &  14:17:27.882 &  -3:58:16.70 &     21.41 & 0.605  \\
41 &  14:17:27.389 &  -3:58:12.06 &     22.42 & ---    \\
42 &  14:17:26.692 &  -4:02:43.83 &     23.98 & 1.199  \\
43 &  14:17:27.001 &  -4:02:15.24 &     23.37 & ---    \\
\enddata
\tablecomments{Column 1: Slit number; Column 2: Right Ascension (J2000); Column 3: Declination (J2000); Column 4: apparent $i$-band magnitude; Column 5: redshift. Slits 3, 26, and 33 are stars denoted with the $\medstar$ symbol. 3C 297 (slit 17) is highlighted in boldface.}
\end{deluxetable}
\end{center}
%---------------------------------------------------------------------

%-----------------------------------------------------------------------------------------------

\section{Analysis of the Optical Spectra}

%-----------------------------------------------------------------------------------------------
\subsection{3C\,297}
%----------------------------------------------------------------------------------------------

%-----------------------------------------------------------------------------------------------
\begin{figure*}
\epsscale{1.2}
\plotone{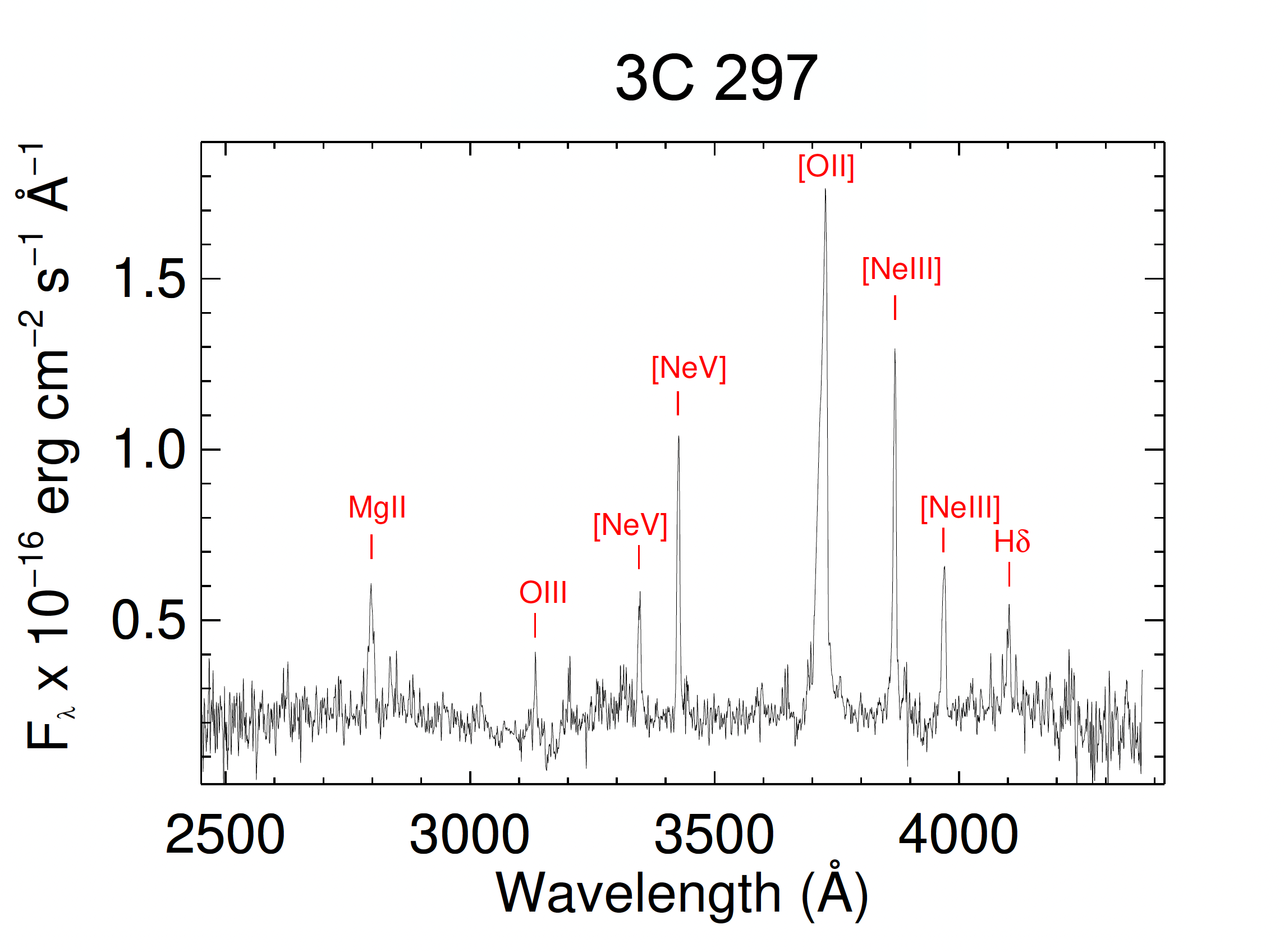}
\caption{New GMOS spectrum of 3C\,297. The spectrum is displayed with a redshift correction of $z$=1.408.
\label{spec3c297}}
\end{figure*}

%-----------------------------------------------------------------------------------------------

Our GMOS spectrum of 3C\,297 is displayed in Fig.\ \ref{spec3c297}. This spectrum has
prominent emission lines: Mg\,{\sc ii}, O\,{\sc iii}, He\,{\sc ii}, [Ne\,{\sc v}], [O\,{\sc ii}], and [Ne\,{\sc iii}]. From the position of the different lines, we measured a mean redshift of $z$=1.408 $\pm$0.001.

Before undertaking our detailed spectral analysis, we applied a redshift correction of $z$=1.408 to the spectrum of 3C\,297. 
We also applied a Galactic extinction correction of $A_{\lambda,V}$=0.146 mag using the extinction maps of \citet{schlafly2011} and the \citet{cardelli1989}
extinction law. In order to increase the
signal-to-noise (S/N) of the emission lines, we smoothed the spectrum using a Gaussian filter of 3 pixels width. This smoothing
preserves the spectral resolution while suppressing high-frequency noise.

The spectrum of 3C\,297 is essentially free of either blue or red continuum emission. Moreover, we do not detect stellar absorption features. 
The presence of Mg\,{\sc ii}~2798~\AA\ and [Ne\,{\sc v}] ~$\lambda\lambda$3345,3424 is further proof of the AGN nature of this object. 
The latter doublet is conspicuous, suggesting that 3C\,279 has high-ionization gas.

The strongest line detected in the observed wavelength interval corresponds to 
[O\,{\sc ii}]~$\lambda\lambda$3726,3729 (hereafter [O\,{\sc ii}]~$\lambda$3728). This line
profile displays a broad, prominent blue asymmetry, indicating the presence of
powerful outflows in the center of the radio galaxy.
In blue-asymmetric oxygen lines a component is usually shifted to the blue by a few hundred km/s or more, relative to the systemic velocity  and is significantly
broader than the narrower component. In some quasars, this blueshifted
line can display FWHM values of a few thousand kilometers per second and the centroid
of the line can be shifted by several hundred kilometers per second relative to the systemic velocity (see \citet {zakamska2014} and discussion below).

%-----------------------------------------------------------------------------------------------
\begin{figure*}
 \epsscale{1.1}
 \plottwo{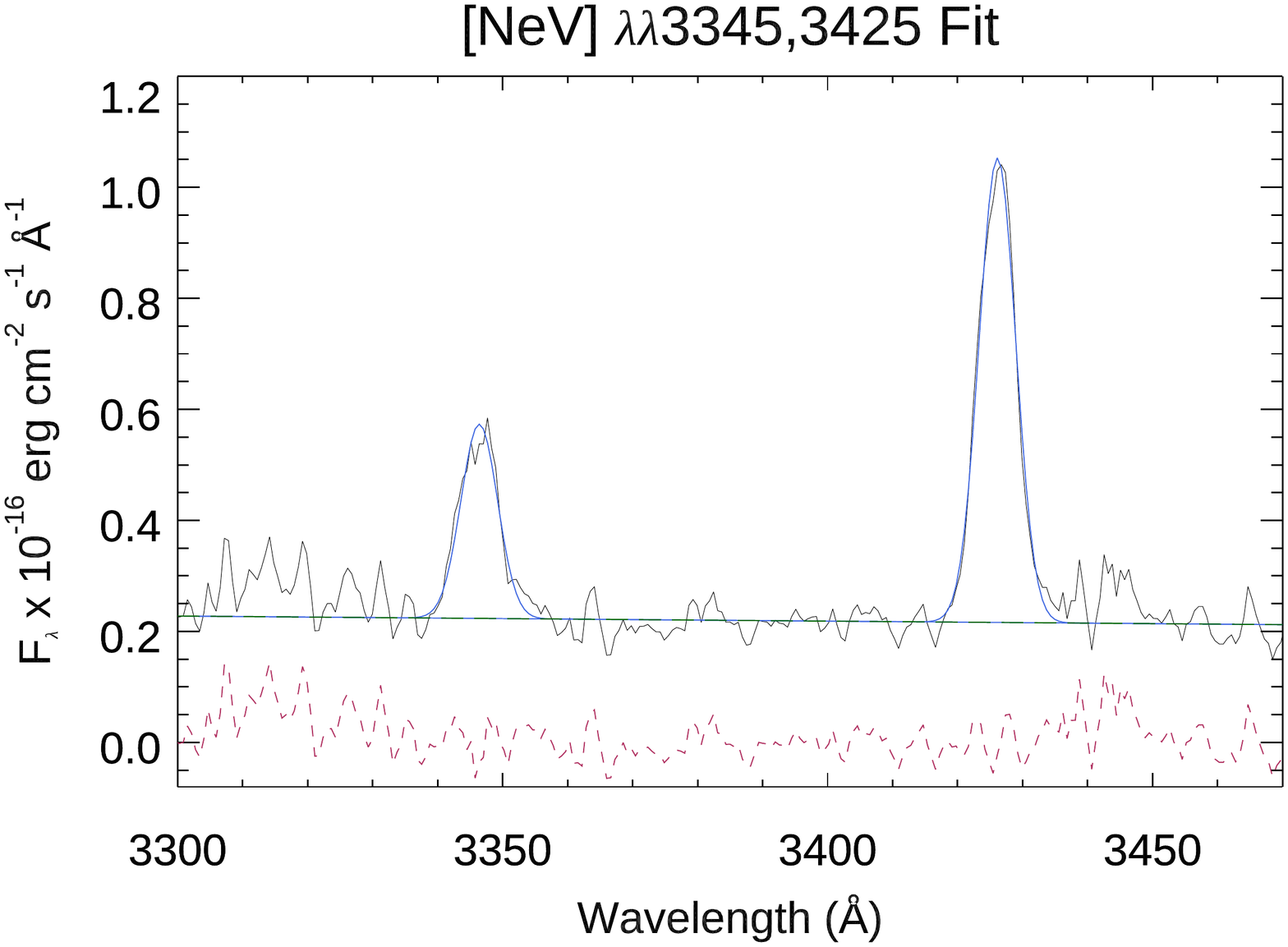}{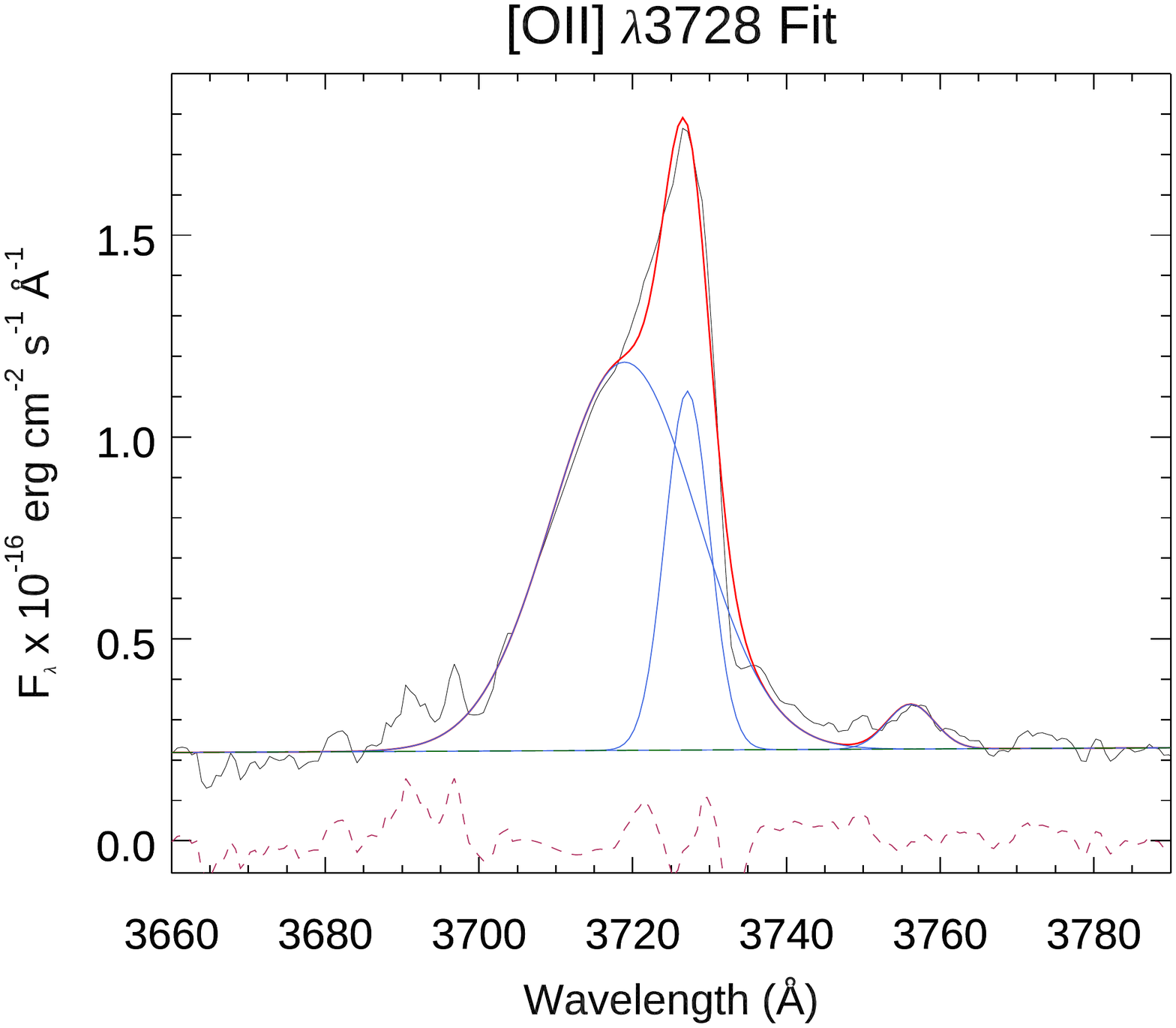}
\caption{Example of the Gaussian decomposition applied to the observed emission line profiles. 
\textit{Left}: Fit to the coronal lines [Ne\,{\sc v}]~$\lambda\lambda$~3345, 3425. \textit{Right}: 
[O\,{\sc ii}]~$\lambda\lambda$~37286, 3729. For this latter line, a second blueshifted, broad component of 
FWHM=1790~km\,s$^{-1}$ was necessary to reproduce the observed profile. In both panels, the 
observed profile is in black, individual Gaussians are in blue, the total fit in red. The green 
line is the continuum level and the dashed maroon line is the residual after subtraction of the fit.
\label{fig:gaussfit}}
\end{figure*}

%-----------------------------------------------------------------------------------------------

In order to measure the flux, centroid position, and FWHM of the emission lines, we fitted Gaussian functions to individual lines, or to sets of blended lines, using  {\sc python} scripts written by our team. Usually, one or two Gaussian components were used to reproduce the observed profile. The criterion for the best solution was the minimum value of the reduced ${\chi}^{2}$.

Typically, one Gaussian is needed to properly fit the observed lines except for [O\,{\sc ii}], where two components were used. Throughout this process, the underlying continuum emission was approximated by a linear fit. Figure~\ref{fig:gaussfit} shows examples of the Gaussian fitting for [Ne\,{\sc v}] and [O\,{\sc ii}]. Table~\ref{tab:fluxes} lists the integrated emission line fluxes of all lines measured at a 3$\sigma$ level, the centroid position, and the FWHM. Notice that all lines except Mg\,{\sc ii} and [O\,{\sc ii}] broad are spectroscopically unresolved (the instrumental FWHM is 6.8~\AA.)

%----------------------------------------------------------------------------------------------

%-----------------------------------------------------------------------------------------------
\begin{table}
\caption{Identifications, integrated fluxes, and FWHM of the lines detected in the spectrum of 3C\,297.}
\label{tab:fluxes}
\centering
\begin{tabular}{lccc}
\hline
Line &	Wavelength &	Flux $\times 10^{-16}$ & FWHM \\
     &  (\AA) & (erg\,cm$^{-2}$\,s$^{-1}$) & (km\,s$^{-1}$) \\
\hline
Mg\,{\sc ii} &	2797.70	& 4.23$\pm$0.49	& 1059 \\
O\,{\sc iii} &	3133.70	& 1.28$\pm$0.13	& 651  \\
{[}Ne\,{\sc v}] &	3346.41	& 2.48$\pm$0.30	& 609 \\
{[}Ne\,{\sc v}] &	3426.13	& 5.92$\pm$0.30	& 595 \\
{[}O\,{\sc ii}]	broad & 3718.94	& 22.73$\pm$0.93  & 1704 \\
{[}O\,{\sc ii}]	& 3727.11	& 6.46$\pm$0.28	& 547 \\
{[}Ne\,{\sc iii}] &	3868.46	& 8.45$\pm$0.39	& 600 \\
{[}Ne\,{\sc iii}]	& 3968.95	& 3.97$\pm$0.39	& 585 \\
H$\delta$	& 4101.34	& 2.38$\pm$0.38	   &  610 \\
\hline
\end{tabular}

\tablecomments{All lines but Mg\,{\sc ii} and [O\,{\sc ii}] broad (see Fig.~\ref{fig:gaussfit} right panel) are spectroscopically unresolved (the instrumental FWHM is 6.8~\AA.)}
\end{table}
%-----------------------------------------------------------------------------------------------

%-----------------------------------------------------------------------------------------------
\subsection{The AGN activity in 3C\,297}
%-----------------------------------------------------------------------------------------------
The continuum emission in 3C\,297 has no hints of either a blue continuum, typical of Type~I AGN,
or stellar absorption features, usually observed in Type~II AGN. However, the detection of coronal lines of
[Ne\,{\sc v}], usually associated with AGN, confirms the presence of an active nucleus in this object (see Figure~\ref{spec3c297}). 

We searched for the presence of broad-line features indicative of a broad-line region (BLR). 
The only permitted lines, detected at a 3$\sigma$ level, are
those of Mg\,{\sc ii}~$\lambda$2798, O\,{\sc iii}~$\lambda$3133 and H$\delta$ at 4101~\AA (see Fig.~\ref{spec3c297}).
The former is significantly broader (FWHM of 1059~km\,s$^{-1}$) than most 
forbidden lines (FWHM $\sim 500-600$~km\,s$^{-1}$) \citep{osterbrock2006}. 
However, Mg\,{\sc ii}~$\lambda$2798 is actually a doublet, ($\lambda 2791$ and $\lambda 2798$), 
and at our spectral resolution the individual components cannot be resolved. With an intrinsic 
FWHM of 600~km\,s$^{-1}$, the Mg\,{\sc ii} doublet is observed as a broad blended profile.

From the lack of continuum features typical of Type~I AGN,  and the absence of broad components in 
the permitted lines, we classify 3C\,297 as a Type~II radio source. 
%-----------------------------------------------------------------------------------------------
\subsection{Outflowing ionized gas}
\label{outflow}
%-----------------------------------------------------------------------------------------------

The broadest feature that we measure in the spectrum of 3C\,297 corresponds to a blueshifted
component found in the forbidden [O\,{\sc ii}]~$\lambda$3728 line. This feature is centered at
3719~\AA, with a FWHM of 1790~km\,s$^{-1}$. In addition, we also detected a narrow component
associated with the same transition, with a FWHM of 547~km\,s$^{-1}$, very similar to the width
of the remaining narrow forbidden lines  (see Figure~\ref{fig:gaussfit}). Because the broad feature
is a forbidden line, its origin is outside a putative Broad Line Region (BLR). The gas density in a BLR region is orders of magnitude larger than the critical density 
of the transition leading to the [O\,{\sc ii}]~$\lambda$3728 doublet. If the broad component is indeed
associated with [O\,{\sc ii}]~$\lambda$3728, it is displaced by 657~km\,s$^{-1}$ to the blue of 
the rest-frame wavelength.

[O\,{\sc iii}]\,$\lambda$5007 falls outside the spectral range of the data. The only published
spectrum covering that line is the one of \citet{jackson1997}, that has low spectral resolution ($R=380$). 
Although the doublet [O\,{\sc iii}]~$\lambda\lambda$4959,5007 is clearly detected, it is not 
possible to confirm from the \citet{jackson1997} data the presence of an obvious blueshifted asymmetric line. 
However, several detections of broad blue-shifted features associated with [O\,{\sc ii}]  have been reported in the
literature. For instance, \citet{balmaverde2016} studied a sample of 224 quasars selected from the Sloan Digital 
Sky Survey (SDSS) at z $<$ 1. They focused on ionized gas outflows traced by the optical [O\,{\sc iii}] and 
[O\,{\sc ii}] lines. Most of the quasar spectra show asymmetries and broad wings in both lines, although the former have 
larger wings than the latter. These line asymmetries in quasars are generally interpreted as signs of outflows instead of inflows from the
farside of the emitting region. Indeed, \citet{perna2015} detected outflows extending to $\sim$10 kpc from 
the central black hole using [O\,{\sc ii}]. \citet{zakamska2014} found that [O\,{\sc ii}]$\lambda\lambda$3726,3729 
also shows outflow signatures, in some cases consistent with extremely broad features seen in [O\,{\sc iii}] 
(see also \citet{davies2015}). We conclude that the strong asymmetry observed in [O\,{\sc ii}] points to an extreme gas outflow in 3C\,297. 
Given the evidence for strong outflows from the spectra, we might expect a disturbed X-ray atmosphere.  Even if the number of counts is very low and therefore any statement should be tentative, in the two areas marked with a dashed red circle and ellipse (see Fig.~\ref{regions}), we highlighted an excess of X-ray counts. In the southwestern region (4\arcsec~circle) this excess has a Gaussian significance of 5.8$\sigma$, while the excess in the northeastern direction (ellipse 4\arcsec $\times$ 2\arcsec)
has a significance of $\sim$3.2$\sigma$.
%-----------------------------------------------------------------------------------------------

%-----------------------------------------------------------------------------------------------
\subsection{Other sources in the field}
%-----------------------------------------------------------------------------------------------

As expected, the spectra of the 39 science targets in the field of 3C\,297 have different signal-to-noise 
ratios (S/N). The spectra for those field sources for which we derived a redshift are presented in the Appendix.

Some objects have bright emission lines that are easy to identify 
(e.g.\ [O\,{\sc iii}] for slits 28 and 40, H$\alpha$ for slits 15 and 31). With only four exceptions discussed 
below, all of our spectra allow the identification of at least two spectral lines, even in those spectra 
with low S/N.

The redshifts of slits 6, 12, 19, and 42 are derived using the [O\,{\sc ii}] line 
only. [O\,{\sc ii}] is the most probable line observed in these spectra for the following reasons. If it were 
[O\,{\sc iii}] ($\lambda 5007$ \AA) we would have detected the [O\,{\sc iii}] ($\lambda 4960$ \AA) line as well. If 
the single line of these four spectra were [NeIII] ($\lambda 3869$ \AA), we would always have [OII] 
and [O\,{\sc iii}] lines too.  

The main conclusion of our field spectroscopy is that we did not identify any galaxies at the same 
redshift as 3C\,297. 

%-----------------------------------------------------------------------------------------------

%-----------------------------------------------------------------------------------------------
\section{Is 3C\,297 a high-$z$ fossil group?}
%-----------------------------------------------------------------------------------------------

Our main findings for 3C\,297, namely the presence of an X-ray-luminous hot gas halo and a lack 
of companion galaxies, would make it a candidate for a so-called fossil group (see, e.g.\ 
\citealt{ponman1994, mendes2007,schirmer2010} and \citealt{aguerri2021} for a recent review). 

Fossil groups, or fossil clusters, are systems where the closest $M^*$ galaxies  have dynamically 
collapsed onto the BCG due to dynamic friction (\citealt{chandra1943}), while the group's common 
X-ray halo would remain visible due to its long cooling time and AGN feedback (\citealt{jones2003}). 
Fossil groups are defined by the presence of an X-ray halo (with $L_X \ge 10^{42}$ erg s$^{-1}$, that is compatible with our estimate, see Sec.~\ref{xray}) 
and a characteristic magnitude gap of $\sim$ 2 mag between their BCG and the second-brightest 
group member \citep{kundert2017,aguerri2021}. 

Observationally, catalogs of fossil groups are usually limited to $z < 0.5$. For instance, 
the highest redshift for the fossil groups reported by a recent survey is $z=0.442$  \citep{johnson2018}. 
In the \citet{hess2012} sample, the highest-redshift fossil group is 0.489. \citet{voevo2010} presented an X-ray-selected sample of X-ray-luminous fossil groups at $z<0.2$. \citet{pratt2016} studied four                                           
very X-ray-luminous fossil groups detected in the Planck Sunyaev-Zeldovich Survey whose temperatures are 5-6 keV with total masses exceeding $10^{14}$M$\astrosun$.                                       

Simulations show that a fossil system may assemble half of its mass in dark matter by redshift $z > 1$, and that the assembled mass at any redshift is generally higher in a fossil groups than in regular groups \citep[][]{dariush2007}. The existence of a fossil group at the redshift of 3C\,297 is therefore compatible with theoretical expectations.

High-redshift examples of fossil groups are very limited with current record holders
standing at $z$=0.47 \citep{yoo2021} and $z$=0.7 \citep{grillo2013}. The fact that we did not find physical companion galaxies makes 3C\,297 one of the highest redshift fossil groups known to date.

\citet{hess2012}, using both
SDSS and VLA data, demonstrated that fossil groups host radio-loud AGN detected at 1.4 GHz. 3C\,297, being a stronger source than the AGN in the sample studied by \citet{hess2012}, shows that fossil 
groups can also host powerful radio sources. 
We want to highlight that, as reported in \citet{lotz2008}, time-scales of galaxy mergers are of the order of 1 Gyr. Even if poorly constrained, 3C\,297 has a cooling time longer than 1 Gyr. 

%-----------------------------------------------------------------------------------------------

%-----------------------------------------------------------------------------------------------
\section{Summary and Conclusions}
%-----------------------------------------------------------------------------------------------%
We have presented an optical/X-ray study of the high-redshift ($z$=1.408) quasar 3C\,297. In the 3C \chn\ Snapshot observation, we observed a complex morphology of the X-ray halo and the detection of the X-ray counterpart of the radio hotspot visible in the VLA image at 8.4 GHz. With new deeper \chn\ observations, we performed the spectral analysis of the bright X-ray hotspot and of the nuclear region. The nucleus is not detected in the radio image. Therefore, we used a tentative position obtained from an available spectral index map.  We obtained an X-ray luminosity of L$_X=2.82^{+0.57}_{-0.46}\times10^{44}$ erg s$^{-1}$ for the hotspot and L$_X=1.15^{+0.53}_{-0.40}\times10^{44}$ erg s$^{-1}$ for the nucleus, in the energy range 0.5-7 keV. We also investigated if the extended X-ray emission, not spherically symmetric, as expected for cluster emission, had a thermal or nonthermal origin, i.e. if it was due to the hot gas or to IC/CMB. We have only been able to put an upper bound on the temperature being 6 keV. Given the distribution of the surrounding X-ray gas, suggesting that 3C\,297 could be the BCG of a galaxy cluster or group, we searched for fainter companions/galaxy members. With new GMOS observations, we have obtained optical spectra of the neighboring sources of our target, and for 19 of them we measured their redshift. The main conclusion of our field spectroscopy is that we did not identify any galaxies at the same redshift as 3C\,297. We therefore postulate that 3C\,297 is a fossil group, being the first at such high redshift. From the optical spectral analysis, we observed that the line profile displays a broad, prominent blue asymmetry, indicating the presence of powerful ionized gas outflows in the center of the radio galaxy.

%-----------------------------------------------------------------------------------------------

%----------------------------------------------------------------------------------------------
\acknowledgments
%----------------------------------------------------------------------------------------------

We thank the anonymous referee for constructive comments on the original manuscript. 
We want to thank W. R. Forman for fruitful discussion and extremely useful comments and suggestions.
This study is based  on observations  obtained at  the Gemini  Observatory  which is operated by AURA under a cooperative  agreement with  the NSF  on behalf  of the Gemini partnership. 
With heartfelt thanks to the Gemini observatory staff that obtained our data and helped us with the data reduction during the challenging conditions of a global pandemic: Germ\'an Gimeno, Kristin Chiboucas, Joan Font-Serra, Jennifer Berghuis, Jos\'e Cortes, Ted Rudyk, 
Lindsay Magill, and Javier Fuentes.

The authors wish to recognize and acknowledge the very significant cultural role and reverence that the summit of Mauna Kea has always had within the indigenous Hawaiian community. We are most fortunate to have the opportunity to obtain astronomical observations from this mountain.

Support for this work was provided  by the National Aeronautics and Space Administration (NASA) through Chandra grants GO1-22112B, GO9-20083X, GO0-21110X and GO1-22087X.
The National Radio Astronomy Observatory is a facility of the National Science Foundation operated under cooperative agreement by Associated Universities, Inc. SAO Image DS9 development has been made possible by funding from the Chandra X-ray Science Center (CXC), the High Energy Astrophysics Science Archive Center (HEASARC) and the JWST Mission office at Space Telescope Science Institute \citep{joye03}.

R.K. acknowledges  support from the Smithsonian Institution and the Chandra High Resolution Camera Project through NASA contract NAS8-03060. 
%W.F. also acknowledges NASA Grants 80NSSC19K0116, GO1-22132X, and GO9-20109X.

A.R.A. acknowledges CNPq (grant 312036/2019-1) for partial support to this work.

C.S. acknowledges support from the MIUR grant FARE ``SMS''.

\software: Astropy \citep{astropy2013}, CIAO \citep{ciao}, Matplotlib \citep{hunter2007}, Numpy \citep{vanderwalt2011}, Sherpa \citep{freeman2001}.

\facilities  Gemini, \chn.

%----------------------------------------------------------------------------------------------

%----------------------------------------------------------------------------------------

\section*{Appendix}

This appendix presents the new Gemini spectra of the 19 galaxies for which we derived a redshift.
%-----------------------------

\begin{figure*}[h]
        \centering
        \begin{tabular}{c} 
        \includegraphics[scale=0.31]{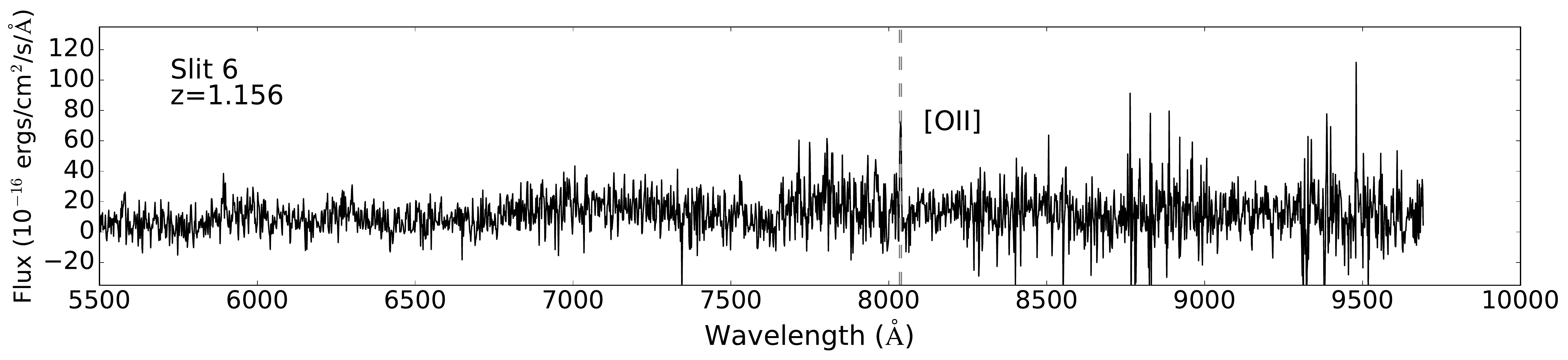} \\    
        \includegraphics[scale=0.31]{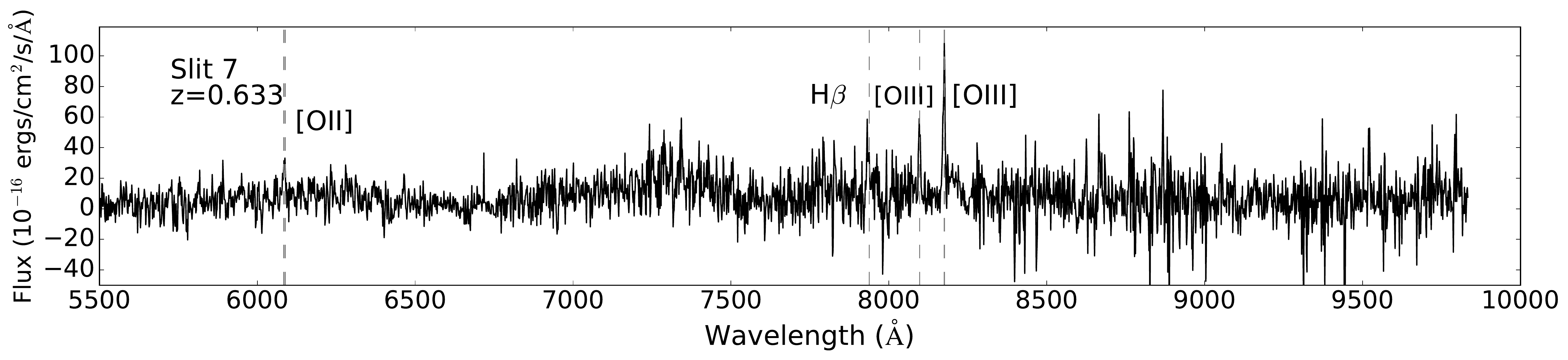} \\ 
        \includegraphics[scale=0.31]{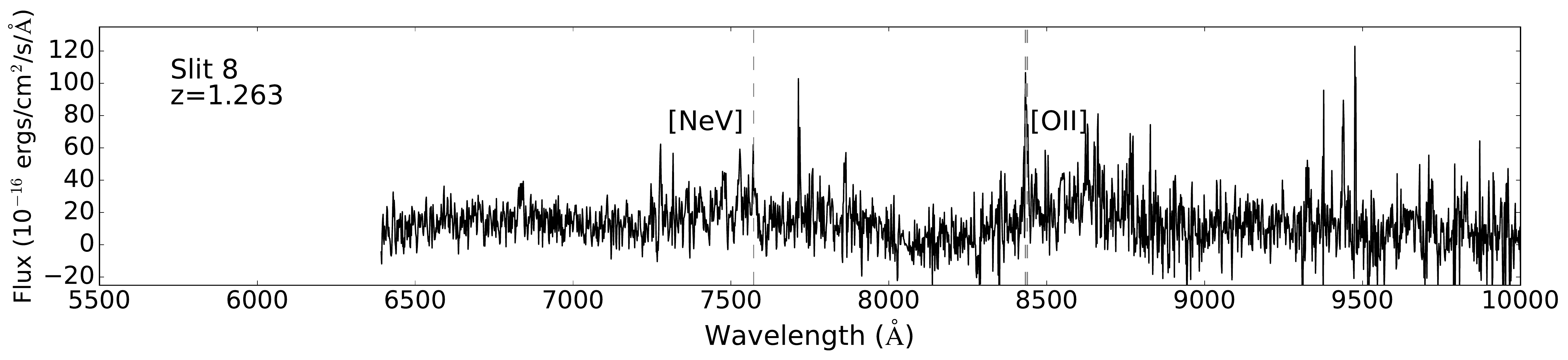} \\
        \includegraphics[scale=0.31]{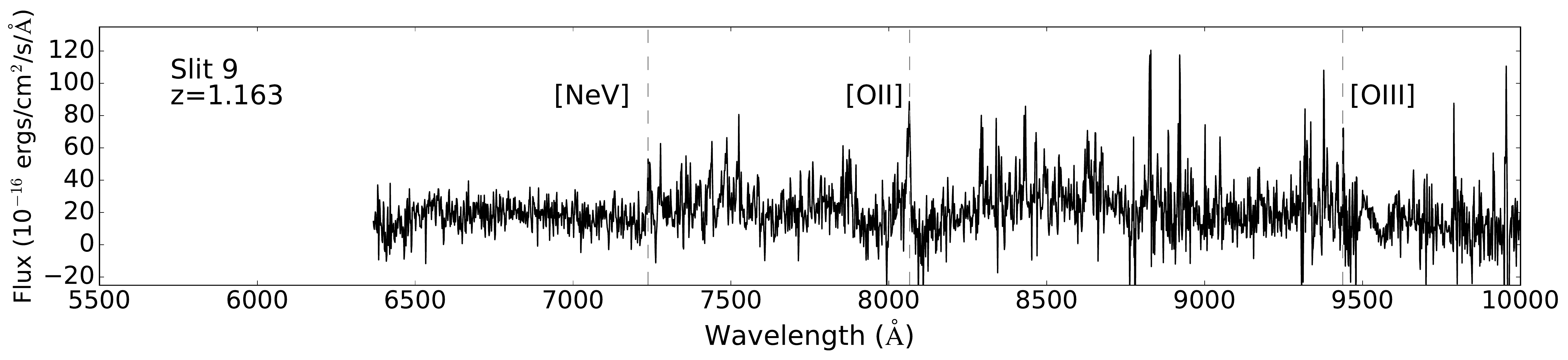} \\ 
        \includegraphics[scale=0.31]{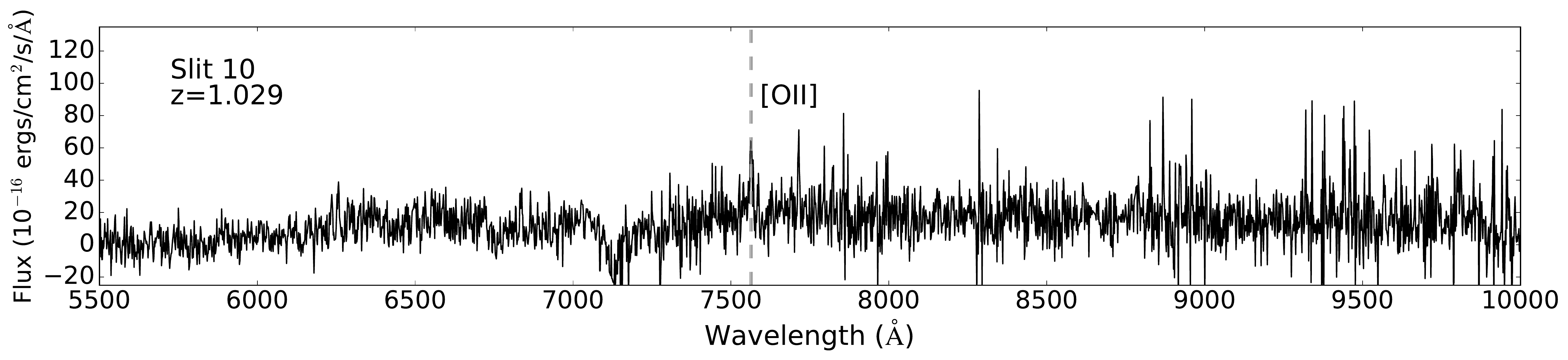} \\  

        \end{tabular}
   \caption{New GMOS spectra of targets in the 3C\,297 field. All spectra are shown in the observed wavelength. The dashed vertical lines mark emission lines that were identified visually. Other apparently significant but non-marked lines are due to noise, which cannot be discerned from real lines at the resolution chosen for these plots.\label{espectros1}}
\end{figure*}
%-----------------------------------------------------------------------------------------------

%-----------------------------------------------------------------------------------------------

\begin{figure*}
        \centering
        \begin{tabular}{c} 
        \includegraphics[scale=0.31]{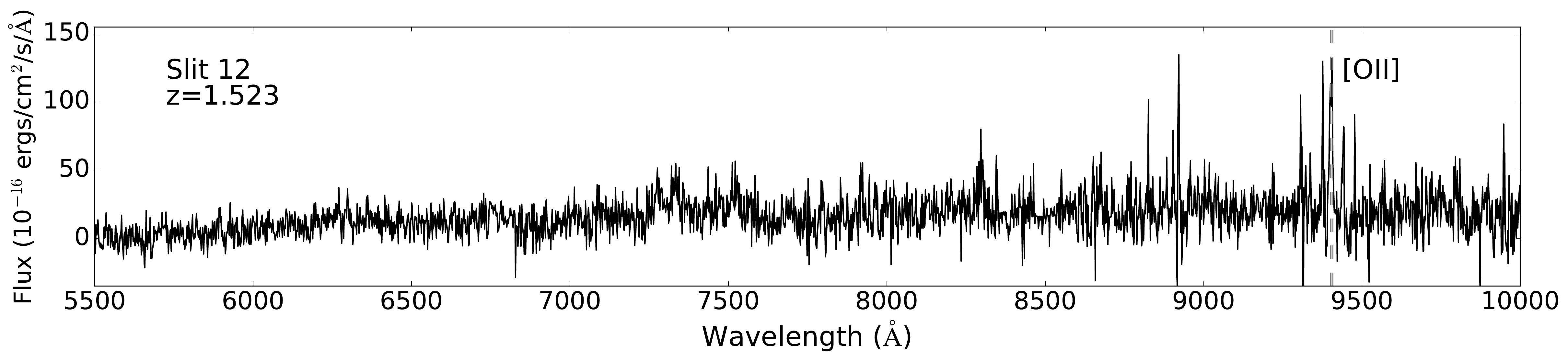} \\
        \includegraphics[scale=0.31]{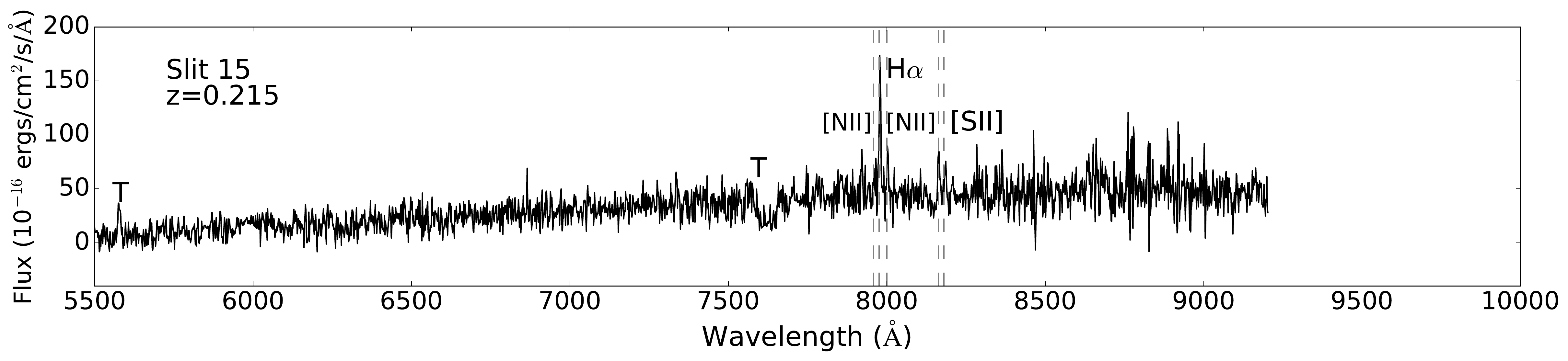} \\ 
        \includegraphics[scale=0.31]{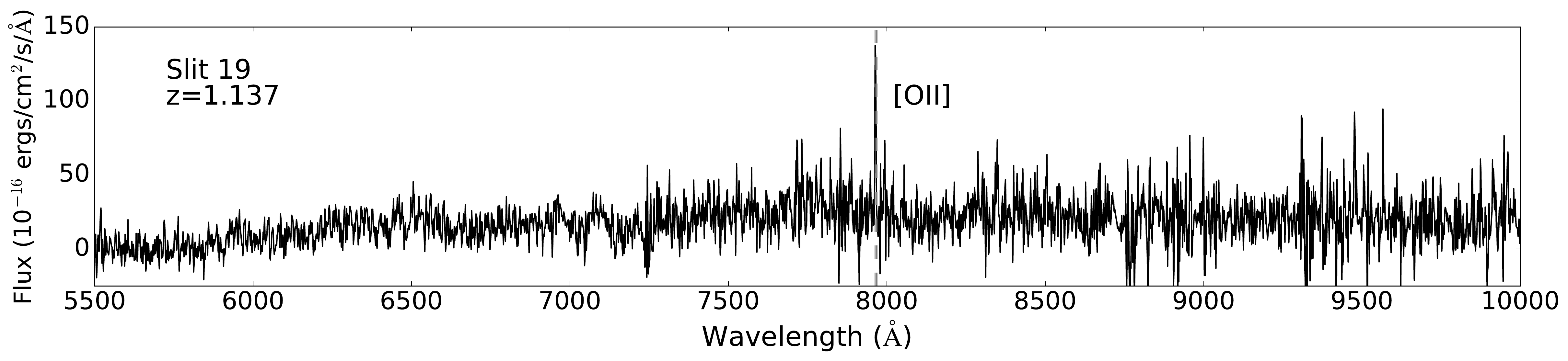} \\ 
        \includegraphics[scale=0.31]{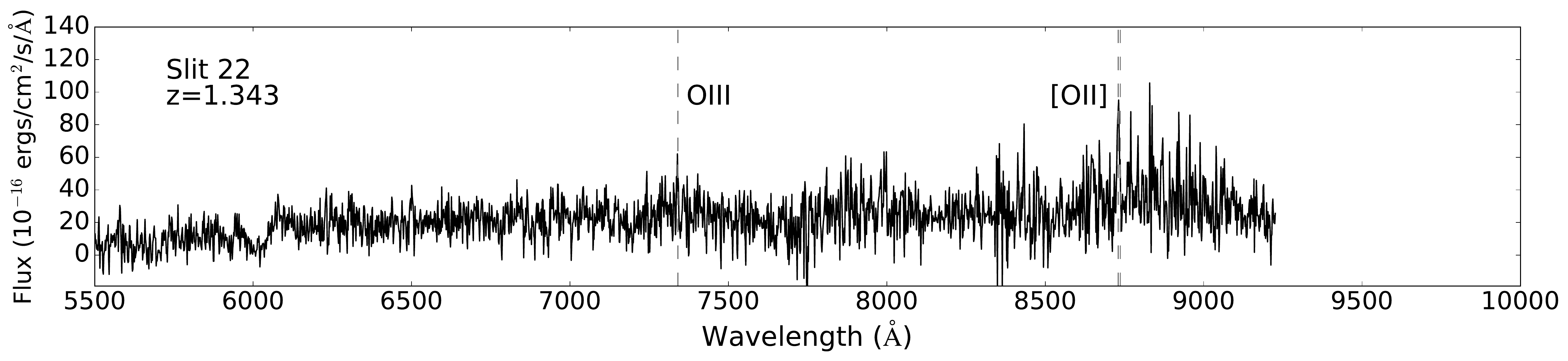} \\
        \includegraphics[scale=0.31]{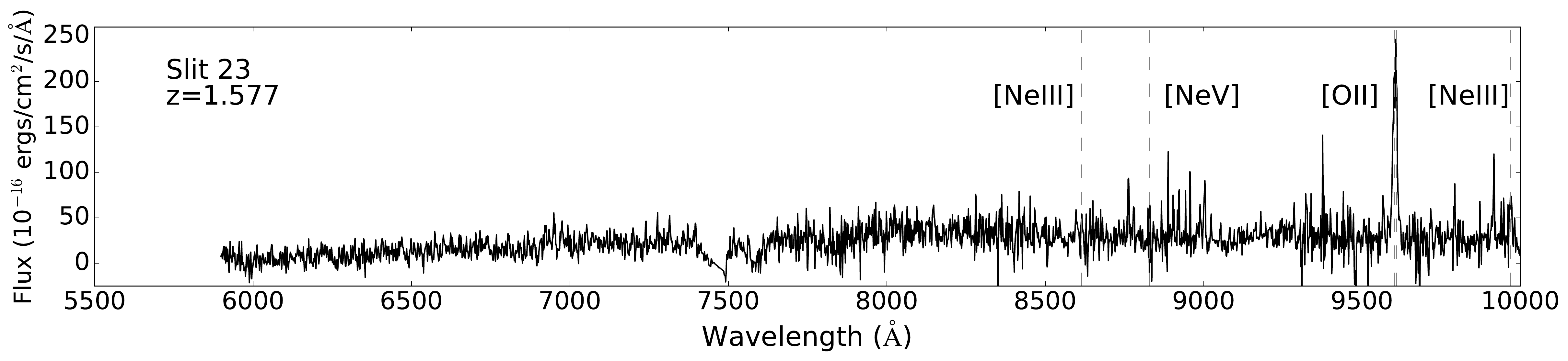} \\
        \includegraphics[scale=0.32]{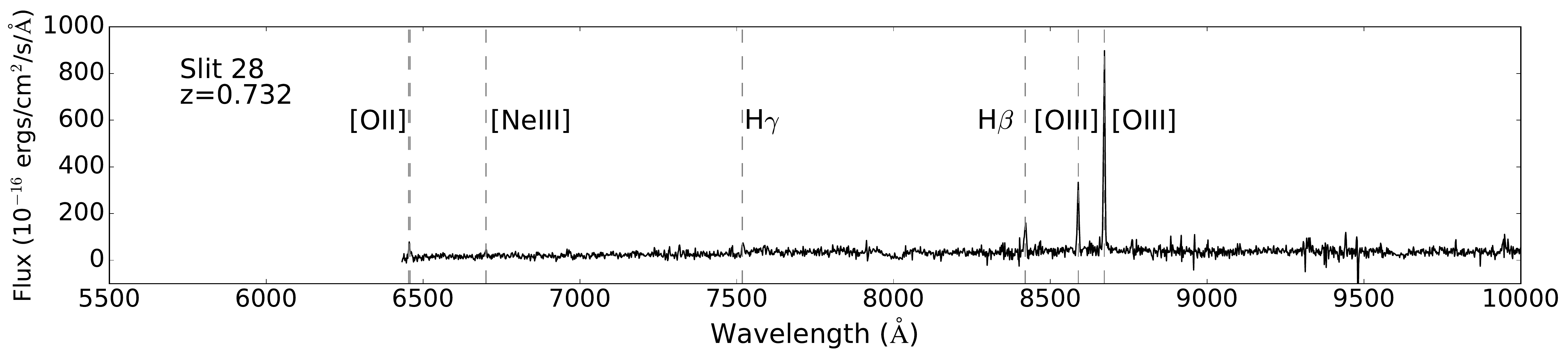} \\
        \end{tabular}
   \caption{New GMOS spectra of targets in the 3C\,297 field. All spectra are shown in the observed wavelength. The dashed vertical lines mark emission lines that were identified visually. Other apparently significant but non-marked lines are due to noise, which cannot be discerned from real lines at the resolution chosen for these plots.\label{espectros2}}
\end{figure*}
%-----------------------------------------------------------------------------------------------

%-----------------------------------------------------------------------------------------------

\begin{figure*}
        \centering
        \begin{tabular}{c} 
        \includegraphics[scale=0.31]{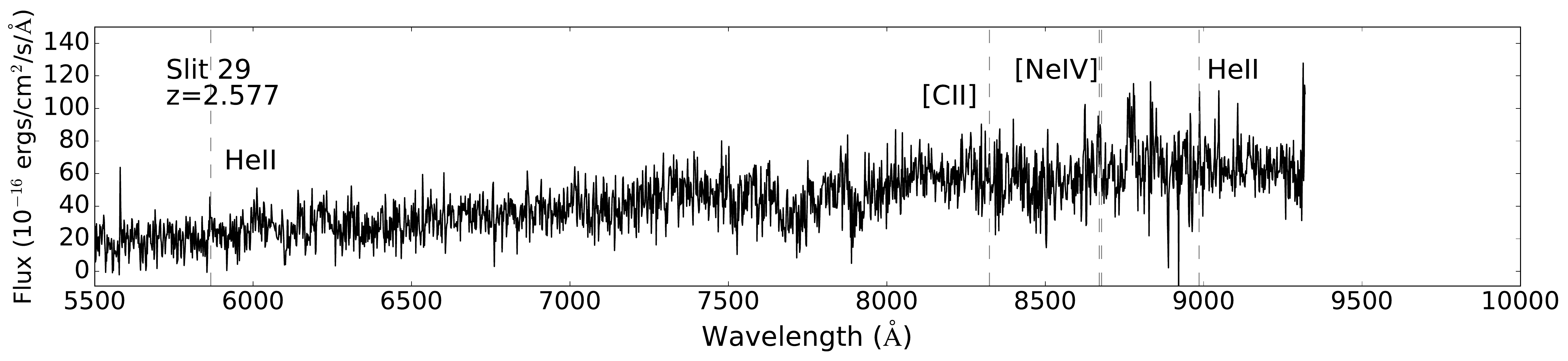} \\     
        \includegraphics[scale=0.31]{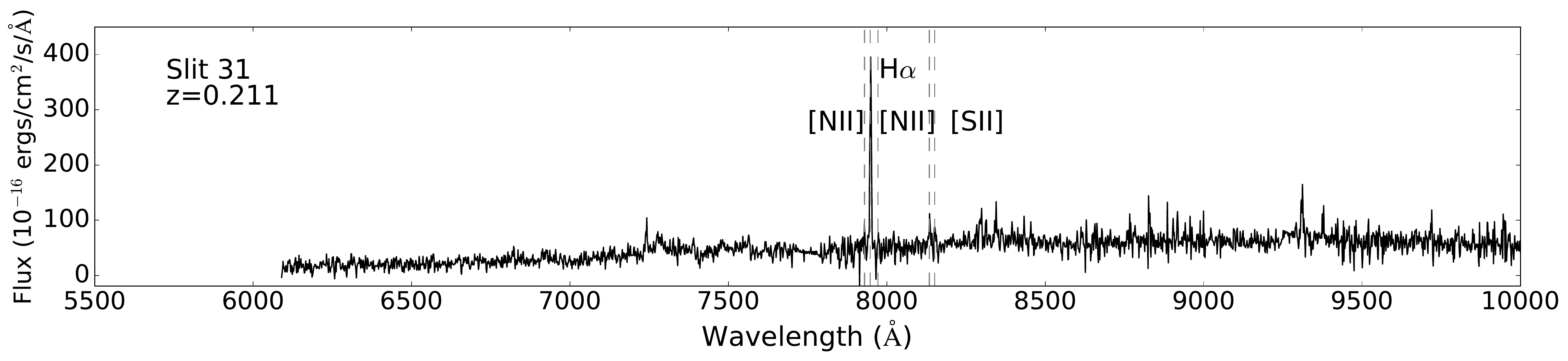} \\ 
        \includegraphics[scale=0.31]{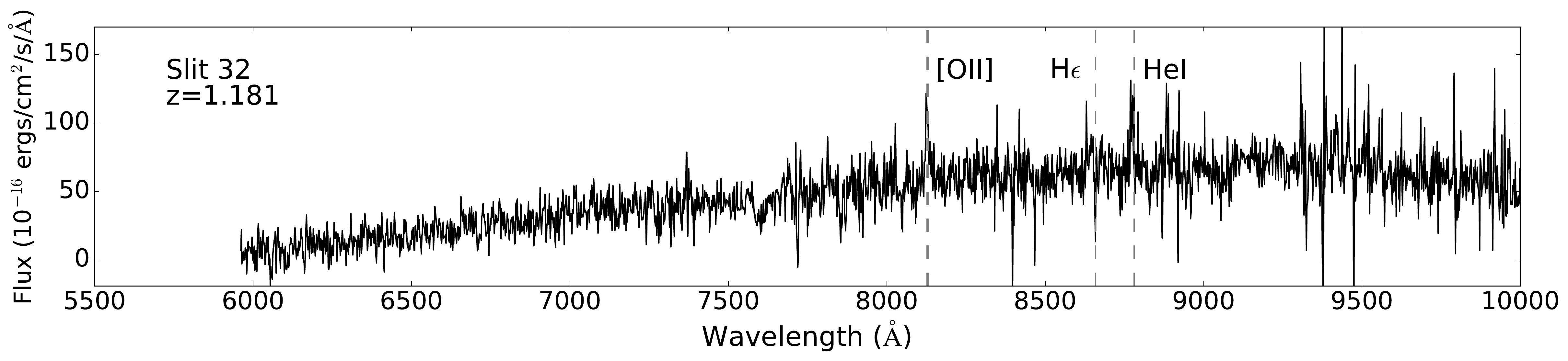} \\      
        \includegraphics[scale=0.31]{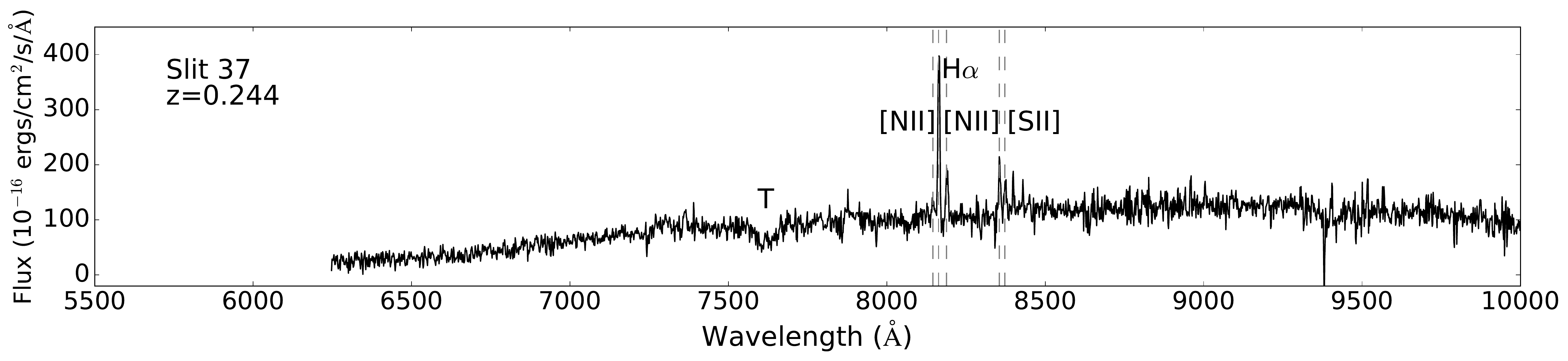} \\
        \includegraphics[scale=0.31]{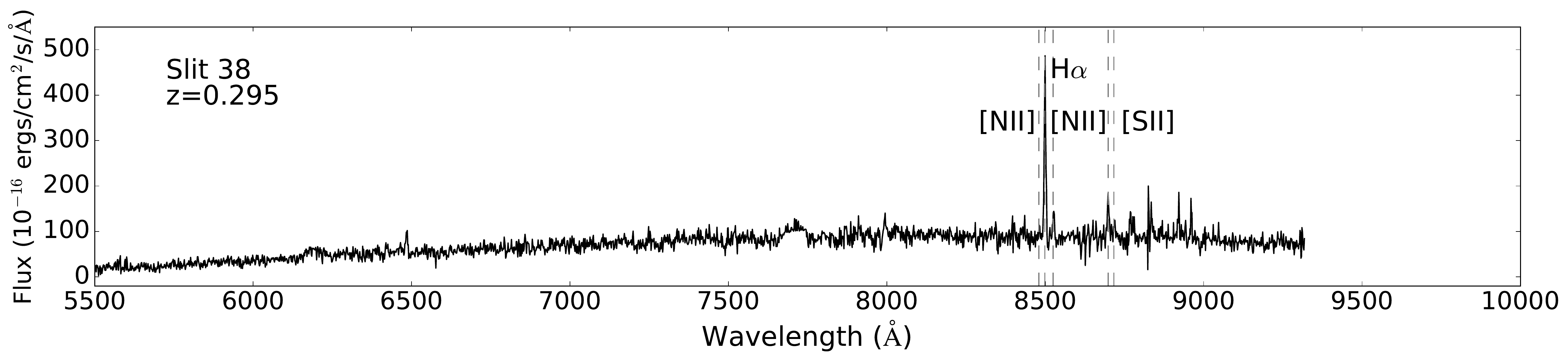} \\
        \includegraphics[scale=0.31]{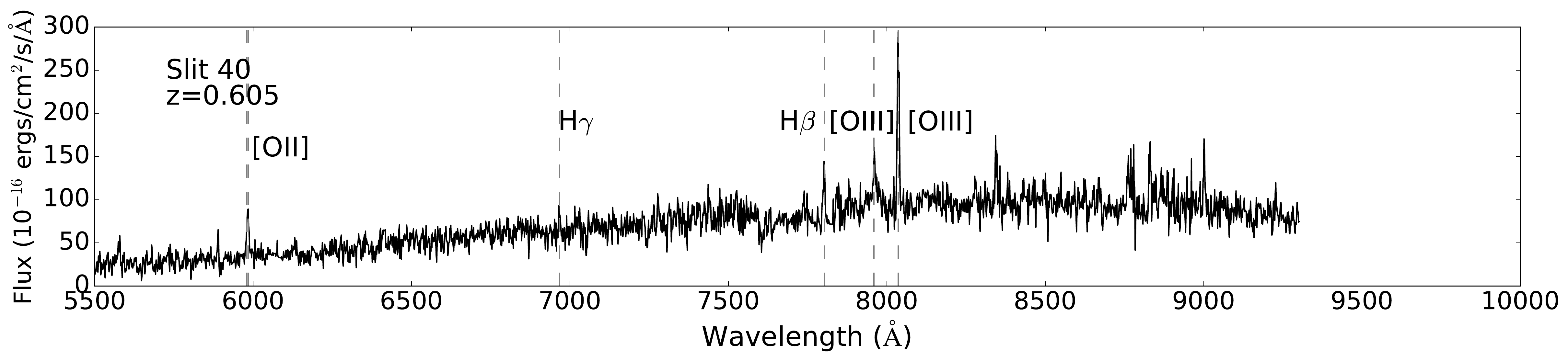} \\
        \end{tabular}
   \caption{New GMOS spectra of targets in the 3C\,297 field. All spectra are shown in the observed wavelength.The dashed vertical lines mark emission lines that were identified visually. Other apparently significant but non-marked lines are due to noise, which cannot be discerned from real lines at the resolution chosen for these plots.\label{espectros3}}
\end{figure*}

%-----------------------------------------------------------------------------------------------
%-----------------------------------------------------------------------------------------------

\begin{figure*}
        \centering
        \begin{tabular}{c} 
        \includegraphics[scale=0.32]{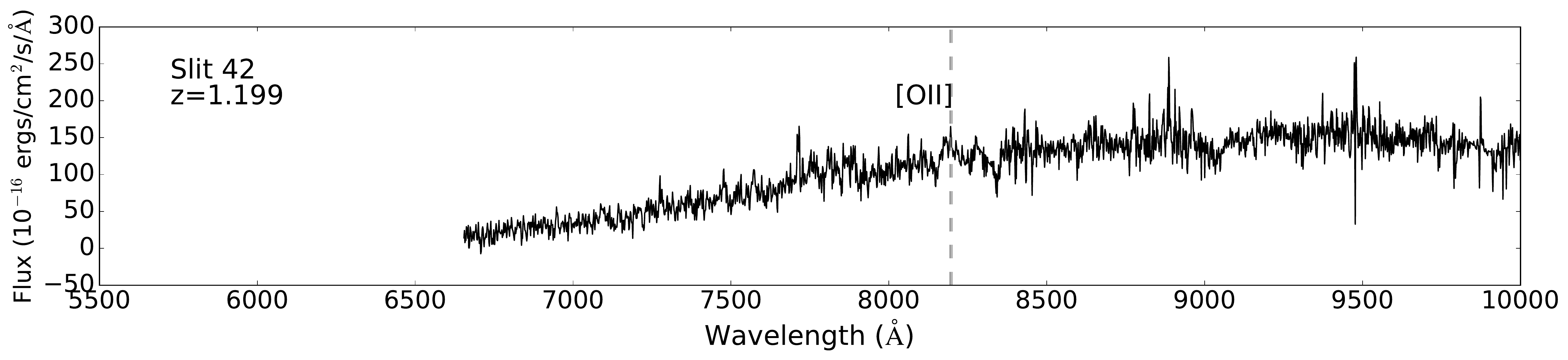}
        \end{tabular}
   \caption{New GMOS spectra of targets in the 3C\,297 field. All spectra are shown in the observed wavelength. The dashed vertical lines mark emission lines that were identified visually. Other apparently significant but non-marked lines are due to noise, which cannot be discerned from real lines at the resolution chosen for these plots.\label{espectros4}}
\end{figure*}

%-----------------------------------------------------------------------------------------------
%-----------------------------------------------------------------------------------------------

%---------------------------------------------------------------------------------------------

%---------------------------------------------------------------------------------------------
\end{document}